\documentclass[aps,pra,reprint,10pt,a4paper,nofootinbib]{revtex4-2}
\usepackage[utf8]{inputenc}
\usepackage{makecell}
\usepackage[sc,osf]{mathpazo}
\usepackage{amsmath}
\usepackage[T1]{fontenc}
\usepackage[margin=0.5in]{geometry}
\usepackage{latexsym}
\usepackage{amssymb}
\usepackage{float}
\usepackage{placeins}
\usepackage[colorlinks=true,citecolor=blue,urlcolor=blue]{hyperref}
\usepackage{color}
\usepackage{graphics,epstopdf}
\usepackage{soul}
\usepackage[demo]{graphicx}
\usepackage{graphicx}
\usepackage{lipsum}
\usepackage{adjustbox}
\usepackage[normalem]{ulem}
\usepackage[table,xcdraw]{xcolor}
\usepackage{braket}
\usepackage{physics}
\usepackage{ragged2e}
\usepackage{hhline} 
\newtheorem{proposition}{Proposition}

\usepackage{comment}
\usepackage{multirow}
\usepackage{array} 
\usepackage{orcidlink}
\usepackage[font=small,labelfont=bf,justification=justified,singlelinecheck=false]{caption}

\newcommand{\aka}[1]{{\color{black}{{#1}}}}
\begin{document}

\title{
Hierarchy of entanglement detection criteria for random  high-dimensional states}

\author{Akhil Kumar Awasthi$^{1, 2}$\, \orcidlink{0009-0000-8971-1420}, Sudipta Mondal$^{1, 2}$\, \orcidlink{0009-0002-3465-9279}, Rivu Gupta$^3$\, \orcidlink{0000-0002-0254-2028},  Aditi Sen(De)$^{1, 2}$\,\orcidlink{0000-0003-1693-0440}}

\affiliation{$^1$  Harish-Chandra Research Institute, Chhatnag Road, Jhunsi, Allahabad - \(211019\), India}

\affiliation{$^2$ Homi Bhabha National Institute, Training School Complex, Anushakti Nagar, Mumbai \(400 094\), India}

\affiliation{$^3$ Dipartimento di Fisica “Aldo Pontremoli,” Università degli Studi di Milano, I-$20133$ Milano, Italy}

\begin{abstract}

Entanglement is a cornerstone in quantum information science, yet detecting it efficiently remains a challenging task. Focusing on states with nonpositive partial transpose (NPT), we establish a hierarchy among entropy-based, majorization, realignment, and reduction criteria for Haar uniformly generated random states in finite dimensions, analyzing their performance based on rank and subsystem dimension. We prove lower bounds on the rank of mixed quantum states beyond which the realignment and entropic criteria fail to detect entanglement.  We evaluate the relative effectiveness of the considered detection methods using three key indicators -- fraction of detected states, mean detectable entanglement, and minimum required entanglement. Our results provide insights into the entanglement thresholds needed for reliable detection, showing that, beyond a certain level of entanglement, all criteria become equally powerful for low-rank states, while a hierarchy emerges among the various criteria for moderate to high ranks.  Intriguingly, the proposed ordering among the considered criteria in qubit-qudit systems is different from that in higher dimensions. Additionally, we establish that the detection efficiency is influenced by the asymmetry in the subsystem dimensions, by illustrating how the realignment criterion behaves more efficiently than other detection methods when the difference between the subsystem dimensions is small. 


\end{abstract}

\maketitle

\section{Introduction}
\label{sec:intro}
Quantum correlations~\cite{discord_prl_ollivier,adesso2016, Bera2018, uola_2020_RMP, Brunner_2014_RMP}, particularly entanglement~\cite{RevModPhys_2009_horo}, having no classical counterpart, are fundamental to the ongoing second revolution in quantum technologies. Entangled states serve as essential resources for achieving quantum advantage in various protocols, including quantum communication~\cite{Bennett_1992_prl, Bennett_1993_prl, Pirandola2015},  quantum cryptography~\cite{Ekert_1991_prl, Gisin_2002_RMP, Scarani_2009_RMP}, and one-way quantum computation~\cite{Raussendorf_2001_prl, Briegel2009}. In these applications, entanglement must often be distributed between two~\cite{Werner1989, Wootters1998_prl, Sents2016} or more parties~\cite{Wei2003_pra, aditi2010_pra, bengtsson_arXiv_2016_multiparty-entanglement, horodecki_arXiv_2024_multipartite-entanglement} to create bipartite or multiparty resource states. Moreover, higher-dimensional systems have been shown to exhibit stronger quantum correlations~\cite{Vaziri_PRL_2002_3d-entanglement, Thew_PRL_2004_time-energy-qutrit-Bell-test, Vertesi_PRL_2010_detection-loophole-qutrit, Ecker_PRX_2019_overcoming-noise-entanglement-distribution, Designolle_PRL_2021_high-d-steering}, which can significantly enhance the performance of quantum information processing tasks~\cite{Correa_PRE_2014, Wang_PRE_2015, Santos_PRE_2019, Dou_EPL_2020, Usui_PRA_2021, Ghosh_PRA_2022, Konar_PRA_2023, Wei_PRL_2019, Nagali_PRL_2010, Bouchard_SA_2017, Lanyon_NP_2009, Babazadeh_PRL_2017, Muralidharan_NJP_2017}. This makes the reliable certification of entanglement in arbitrary dimensions a critical task.

Over the past three decades, numerous entanglement detection techniques have been developed~\cite{Terhal2002, Hofman_PRA_2003_uncertainty-relation-entanglement, Yu_PRL_2005_local-orthogonal-observable-entanglement, deVicente_QIC_2007_Bloch-representation-entanglement, deVicente_JPA_2008_correlation-matrix-entanglement, Zhang_PRA_2008_stronger-than-realignment, Guhne_PR_2009_entanglement-detection, Shapourian_PRX_2021_diagrammatic-entanglement}. Some are theoretically elegant, some experimentally accessible, and others possess direct operational significance. These include Bell inequalities~\cite{Bell_Physics_1964_Bell-inequality, Clauser_PRL_1969_CHSH, Terhal_PLA_2000_Bell-inequality-separability}, partial transposition (PT)~\cite{peres_prl_1996_ppt, HORODECKI19961}, entropic~\cite{ent_horo_1996}, reduction~\cite{Reduction_horo_1999, Cerf_PRA_1999_reduction}, majorization~\cite{maj_nei_1999, Nielsen_PRL_2001_majorization}, and realignment ~\cite{chen2003matrixrealignment, Rudolph2000_jop, realighment_rudo_2003, Rudolph2005} criteria, entanglement witnesses~\cite{Terhal_LAA_2001_indecomposable-positive-maps, Guhne_JMO_2003_witness, Chruciski_JPA_2014_witness}, covariance matrix approaches~\cite{Guhne_PRL_2007_covariance-matrix, Gittsovich_PRA_2008_covariance-matrix}, and semidefinite programming-based methods~\cite{Bruss_JMO_2002_Innsbruck-Hannover-program-detection, Doherty_PRA_2004_semidefinite-program-detection, Hulpke_JPA_2005_algorithm-detection}. Among these, PT provides a necessary and sufficient condition  for qubit-qubit and qubit-qutrit systems; in higher dimensions, it offers a sufficient condition and fails to detect bound entangled states with positive partial transpose (PPT). The reduction, realignment, majorization, and entropic criteria provide only sufficient conditions for detecting entanglement, with the realignment criterion notably capable of identifying certain bound entangled states.
While pure-state entanglement can be characterized by the von-Neumann entropy of reduced density matrices, mixed-state quantification relies on computable measures like logarithmic negativity \cite{Vidal_PRA_2002_negativity, Plenio_PRL_2005_log-neg} in arbitrary dimensions, and entanglement of formation for two-qubits \cite{Wootters1998_prl}. Given the difficulty of preparing pure states and their limited utility in noisy environments, it, therefore, becomes imperative to determine the efficacy of existing detection criteria in successfully predicting states to be entangled in finite dimensions, especially beyond two-qubit and qubit-qutrit systems.

Random quantum states generated according to the Haar measure (the unique unitarily invariant measure over the unitary group) provide a statistically unbiased way to explore the vast landscape of quantum characteristics in quantum states~\cite{Kendon2002_pra, BengtssonZyczkowski2017, Hayden, Enriquez, Pandit, Gross, Soorya, Banerjee_PRA_2020_decoherence-localisable-entanglement, Hastings}. Studying random states can be important for benchmarking quantum algorithms~\cite{Wallman_NJP_2014_randomized-benchmarking, Epstein_PRA_2014_limits-randomized-benchmarking, Granade_NJP_2015_accelerated-randomized-benchmarking, Alexander_PRA_2016_randomized-benchmarking-MBQC}, understanding typical properties of entanglement~\cite{Znidari_JPA_2007_random-entanglement-witness, Hamma_PRL_2012_entanglement-random-physical, Gupta_PRA_2022_non-Markovianity, Gupta_PRA_2022_catalysis, Iannotti_Quantum_2025_entanglement-stabilizer-entropy-random}, analyzing random quantum circuits~\cite{DeCross_PRX_2025_random-circuit-computational-power, Lee_arXiv_2025_random-circuit-simulation}, and evaluating the performance of quantum protocols on average~\cite{Gupta_PRA_2021_dense-coding-teleportation-random, Muhuri_PRA_2024_dense-coding-non-Markovian}. Additionally, these random states naturally arise in scenarios involving quantum chaos and decoherence~\cite{Haake}, making them valuable for probing the statistical nature of entanglement, coherence, and other quantum correlations. Moreover, most existing studies focus primarily on multiqubit random states, with limited attention given to multiqudit systems, although studies in higher local dimensions are essential to exhibit richer entanglement structures and enhanced quantum advantage.

In this work, we systematically analyze four widely-used entanglement detection criteria, namely majorization, reduction, realignment, and entropic, for random bipartite quantum states by varying dimensions as well as ranks, and compare their performances relative to the PT criterion. We analytically prove lower bounds on the rank of two-qudit states, expressed in terms of their local dimensions, above which both the entropy and realignment criteria fail to detect entanglement, thereby highlighting their limitations, particularly for high-rank mixed states. By generating nonpositive partially transposed (NPT) states, we show how detection efficiency decreases with increasing rank, indicating that these criteria are generally less capable of identifying weakly entangled states. Interestingly, we observe a rank-dependent reversal in performance between majorization and realignment -- while majorization is more effective in qubit-qudit systems, realignment overtakes it in higher dimensions with low ranks. Additionally, we present an independent proof of the equivalence between the PT and the reduction criteria for qubit-qudit systems.

We further explore two key questions for randomly generated higher-dimensional states --  $(1)$ Is there a threshold entanglement content necessary for a criterion to operate effectively? $(2)$ What role does the local dimension play in the performance of detection methods? To address the first question, we introduce the mean and minimum entanglement detectable by a given criterion among simulated NPT entangled states. While the former sheds light on the average value of entanglement required for a certain criterion to be effective for a given rank and dimension, the latter can be used to judge how powerful it is. Specifically, a lower minimum detectable entanglement hints at a superior criterion, implying that it can detect states even with meager entanglement content. Furthermore, the mean and minimum entanglement provide bounds on the resource necessary for states to be successfully detected. While all criteria are equally adept at detecting states of rank two, for full-rank states, most become extremely weak. The hierarchy established through the fraction of detected states remains intact with respect to the other two figures of merit. To address the second question, we uncover a nontrivial relationship between detection efficacy and asymmetry in subsystem dimensions. Specifically, we find that the realignment criterion remains surprisingly effective at high ranks when the two subsystems have nearly equal dimensions, highlighting the crucial role played not only by total dimension and rank but also by the internal structure of the bipartite system.



The remainder of the paper is organized as follows. In Sec.~\ref{sec:pre}, we recall the definitions of the entanglement measures and elaborate on their existing inter-relations. In Sec.~\ref{sec:det_frac_state}, the lower bounds on the ranks of quantum states for entropic and realignment criteria to function are proven, and the fraction of detected states is calculated to establish a preliminary hierarchy among the criteria. Bounds on the mean and minimum entanglement necessary for successful detection are presented in Sec.~\ref {sec:bounds_ent_success}, where the hierarchy is further re-established. In \ref{subsec:mean_min_same_dimension}, we demonstrate that asymmetry in the local dimension can play a role in the effectiveness of the detection criteria. We end our work with discussions and open questions in Sec.~\ref{sec:conclusion}.


\section{Definition and relations between entanglement detection criteria}
\label{sec:pre}

An arbitrary bipartite density matrix, $\rho_{12}$, acting on a composite complex Hilbert space, $\mathbb{C}^{d_1} \otimes \mathbb{C}^{d_2}$, hereafter abbreviated as \(d_1 \otimes d_2\), can be represented as 
\begin{equation}
    \rho_{12} = \sum_{1\leq i,j\leq d_1} \sum_{1\leq \mu,\nu\leq d_2} a^{\mu\nu}_{ij} (|i\rangle\langle j|)_1 \otimes (|\mu\rangle\langle \nu|)_2,
    \label{eq:state}
\end{equation}
where $\{ |i (j)\rangle \}$, and $\{ |\mu (\nu)\rangle \}$ are the sets of  orthonormal vectors in $\mathbb{C}^{d_1}$ and $\mathbb{C}^{d_2}$ respectively, with ${d_1} = \dim \mathbb{C}^{d_1}$ and ${d_2} = \dim \mathbb{C}^{d_2}$. Dating back to the early twentieth century~\cite{Einstein_PR_1935_EPR}, a plethora of criteria have been developed for detecting and quantifying the entanglement inherent in quantum states~\cite{Guhne_PR_2009_entanglement-detection}. While some of these criteria can accurately separate between entangled and separable states, especially in low dimensions, others perform exceptionally well in specific scenarios but are less effective depending on the nature of the quantum system under consideration. We are interested in assessing the effectiveness of these entanglement criteria for typical states in arbitrary dimensions. Towards this aim, we focus on a selection of the most widely recognized and frequently used entanglement detection criteria in the literature. These include the \textit{ partial transposition}, \(\mathcal{P}\)~\cite{peres_prl_1996_ppt, HORODECKI19961}, \textit{reduction}, $\mathrm{R}_D$~\cite{Reduction_horo_1999, Cerf_PRA_1999_reduction}, \textit{majorization}, $\mathcal{M}$~\cite{maj_nei_1999}, \textit{realignment}, $\mathrm{R}_L$~\cite{chen2003matrixrealignment,realighment_rudo_2003}, and \textit{entropy}, $\mathrm{E}$~\cite{ent_horo_1996} criteria. For convenience, we denote the set of these criteria as

\begin{eqnarray}
\mathcal{S} = \{ \mathcal{P},~\text{\textcolor{black}{$\mathrm{R}_D$}},~\mathcal{M},~\mathrm{E},~\text{\textcolor{black}{$\mathrm{R}_L$}} \}.
\label{eq:measure_set}
\end{eqnarray}
 Let us briefly elucidate the considered detection criteria.

     \textit{Partial transposition criterion} ($\mathcal{P}$). The partial transposition, $\rho^{T_{1}}_{12}$, of a density matrix $\rho_{12}$ with respect to the subsystem $1$, is defined as
        \begin{equation}
            \rho^{T_1}_{12} = \sum_{1\leq i,j\leq d_1} \sum_{1\leq \mu,\nu\leq d_2} a^{\mu\nu}_{ij} (|j\rangle\langle i|)_1 \otimes (|\mu\rangle\langle \nu|)_2.
            \label{eq:PPT}
        \end{equation}
    If a state $\rho_{12}$ is separable, then $\rho^{T_1}_{12} \geq 0$ and similarly for the partial transposition with respect to the subsystem $2$. Here, $\sigma \geq 0$ means that $\sigma$ has non-negative eigenvalues (i.e., it is positive semidefinite). States with non-positive partial transpose (NPT) possess entanglement which is distillable (convertible via local operations and classical communications on a large number of copies to a smaller number of maximally entangled states~\cite{Bennett_PRL_1996_distillation}) in $2\otimes d$ while the same is not guaranteed in higher dimensions~\cite{Horodecki_PRL_NPT-distillable}.

 \textit{Reduction criterion} ($\mathrm{R}_D$).  \aka{If $\rho_{12}$ is separable}, then
        \begin{equation}
            \rho_1 \otimes I_2 - \rho_{12} \geq 0 \quad \text{and} \quad I_1 \otimes \rho_2 - \rho_{12} \geq 0,
            \label{eq:reduction}
        \end{equation}
    otherwise the state is entangled~\cite{Reduction_horo_1999, Cerf_PRA_1999_reduction}. $I_{1(2)}$ denotes the $d_{1(2)}$-dimensional identity matrix, whereas $\rho_{1(2)} = \Tr_{2(1)} \rho_{12}$ is the reduced density matrix corresponding to subsystem $1(2)$.

 \textit{Entropy criterion} ($\mathrm{E}$). If a state \(\rho_{12}\) is separable, then  
        \begin{equation}
            S(\rho_{12}) \geq S(\rho_1), \quad S(\rho_{12}) \geq S(\rho_2),
            \label{eq:entropy}
        \end{equation}
    otherwise it is entangled~\cite{ent_horo_1996}. The von Neumann entropy, $S(\sigma)$ is defined as $S(\sigma) = - \Tr \sigma \log_2 \sigma$ for a quantum state, $\sigma$~\cite{nielsen_2010, Preskill, wilde_2013, Watrous_2018}.

\textit{Majorization criterion} ($\mathcal{M}$). For a separable state, $\rho_{12}$, the following inequalities hold:
    \begin{equation}
        \lambda(\rho_{12}) \prec \lambda(\rho_1), \quad \lambda(\rho_{12}) \prec \lambda(\rho_2),
        \label{eq:majorization}
    \end{equation}
    where $\lambda(\sigma)$ is the set of eigenvalues of the denisty matrix, $\sigma$, arranged in descending order, and $ x \prec y \implies \sum_{i=1}^{k} x_i \leq \sum_{i=1}^{k} y_i ~\forall~ k = 1, 2, \dots, l (m)$, for two lists $x$, and $y$, containing $l$ and $m$ elements respectively arranged in descending order. For $l < m$, the former list is appended with $m - l$ zeros at the end, and vice versa~\cite{Nielsen_PRL_2001_majorization, maj_nei_1999}.

\textit{Realignment criterion} ($\mathrm{R}_L$). For a state, $\rho_{12}$, its realignment matrix, $\mathcal{G}$, is defined as
        \begin{equation}
            \rho^{\mathrm{R}_L}_{12} = \sum_{k,l} \mathcal{G}_{kl} \tilde{G}_k^{1} \otimes \tilde{G}_{l}^{2},
            \label{realignment}
        \end{equation}
    with $1 \leq k (l) \leq d_{1(2)}^2$, and $\tilde{G}_k^{1(2)} = \{|i(\mu)\rangle \langle j(\nu)|\}$ are complete sets of orthonormal Hilbert-Schmidt projectors acting on the respective Hilbert spaces. Let $\{g_i\}_{i = 1}^{\min[d^2_1, d^2_2]}$ be the set of singular values of $\mathcal{G}$. If $\rho_{12}$ is separable, then $\sum_i g_i \leq 1$~\cite{chen2003matrixrealignment, Rudolph2000_jop}.

\subsection{Existing relations between the criteria}
\label{subsec:existing_relation}
Our aim is to investigate the interconnections among the elements of the set $\mathcal{S}$ defined in Eq.~(\ref{eq:measure_set}) is guided by prior foundational results on the relationships between these criteria~\cite{Terhal2002}. For the $2 \otimes 2$ and $2 \otimes 3$ systems, the PPT criterion, $\mathcal{P}$, is a necessary and sufficient condition for separability~\cite{HORODECKI19961}. However, beyond these low-dimensional cases, $\mathcal{P}$ remains only a necessary condition for the same~\cite{HORODECKI1997333}, there existing entangled states, known as bound entangled states, which satisfy $\mathcal{P}$~\cite{HORODECKI1997333, Bennett_PRL_1999_UPB-bound-entanglement, DiVincenzo_CMP_2003_UPB-bound-entanglement}.
Since the reduction criterion, $\mathrm{R}_D$, is equivalent to $\mathcal{P}$ in $2 \otimes d$~\cite{JBatle_2004_iop_allconnnection}, it is also necessary and sufficient for separability in $2 \otimes 2$ and $2 \otimes 3$. Violation of the reduction criterion, $\mathrm{R}_D$, implies distillability of entanglement~\cite{Reduction_horo_1999, Cerf_PRA_1999_reduction}, i.e., states violating $\mathrm{R}_D$ definitely violate $\mathcal{P}$, but not the other way around. In other words, \(\mathrm{R}_D\) is generally weaker than \(\mathcal{P}\) in detecting entanglement beyond $2 \otimes d$, for example, in the case of \(3\otimes d\) systems, states that satisfy the reduction criterion are significantly more prevalent than those that meet the PPT criterion~\cite{JBatle_2004_iop_allconnnection}. In the bipartite setting, the majorization criterion, $\mathcal{M}$, is weaker than \(\mathrm{R}_D\) in detecting entanglement, but together they impose a necessary condition for the distillability of a quantum state~\cite{maj_red_2003}. The correspondence between the \(\mathcal{M}\) and \(\mathcal{P}\) criterion becomes progressively weaker as the subsystem dimension \(d\) increases~\cite{JBatle_2004_iop_allconnnection}.

In contrast, the entropy criterion, $\mathrm{E}$, ranks as the weakest in $\mathcal{S}$ due to its reliance on global state properties rather than operational separability tests~\cite{RevModPhys_2009_horo, Bengtsson2017}. There exist some numerical relations between \(\mathcal{P}\), $\mathrm{R}_D$, and \(\mathrm{E}\), suggesting that they behave similarly in \(2\otimes d\) and \(3\otimes d\) ~\cite{JBatle_2004_iop_allconnnection}. For dimensions higher than \(3 \otimes d\), \(\mathcal{M}\) and \(\alpha\)-entropic criteria \(\mathrm{E}_{\alpha}\) yield nearly identical results, indicating a strong agreement between the two in this setting~\cite{JBatle_2004_iop_allconnnection}. Note that, due to the Schur-concavity of the von Neumann entropy~\cite{Kvalseth_IEEE_2022_Schur-concavity-entropy}, states detected as entangled by $\mathrm{E}$ are also detected by $\mathcal{M}$~\cite{Wehrl_RMP_1978_entropy, Marshall_book_2011_majorization}. The interrelation between positive maps, weak majorization, and entropic inequalities plays a significant role in the detection of entanglement in bipartite quantum systems~\cite{Augusiak_2009_maj_ent}.
  
Finally, the realignment criterion $\mathrm{R}_L$, while slightly weaker than $\mathcal{P}$ in low-dimensional systems~\cite{chen2003matrixrealignment,realighment_rudo_2003}, exhibits unique advantages in higher dimensions. Notably, $\mathrm{R}_L$ can detect certain bound entangled states that evade $\mathcal{P}$~\cite{Horodecki_1998,Rudolph2005}. It has been related to the Schmidt coefficients of a bipartite density matrix~\cite{Lupo_JPA_2008_realignment-schmidt}, which on their own form a criterion for entanglement detection in pure states~\cite{SenDe_arXiv_2005_separability-vs-entanglement, nielsen_2010, Preskill, wilde_2013, Watrous_2018}. Inequalities have been conjectured as necessary conditions for bipartite separability based on it, but independent of the parent criterion, and numerically verified for $2 \otimes 3$ systems~\cite{Lupo_JPA_2008_realignment-schmidt}. We will also show that $\mathrm{R}_L$ behaves weakly for high rank states compared to $\mathrm{R}_D$.

\section{Effectiveness of various detection criteria: Role of rank
}
\label{sec:det_frac_state}

While all qualitative criteria for bipartite entanglement detection are equivalent for pure states, their ability to detect mixed state entanglement is widely different. Our main objective is to analyze {\it how well} a considered detection criterion is able to conclude that a state with distillable entanglement is actually entangled. For this purpose, we resort to the PPT criterion as a benchmark to first determine whether a state possesses distillable entanglement. If so, we then apply the set of chosen criteria, given in Eq.~\eqref{eq:measure_set}, to the state and check which of them can detect the state as entangled. This depends, to a large extent, on the subsystem dimensions and the rank of the mixed state; generally, entanglement detection becomes more robust as the total dimension increases, but more challenging as the rank grows. While previous studies were often restricted to low-dimensional systems or specific classes of states, we systematically analyze the variation in detection ability across the full range of dimensions and ranks by simulating bipartite density matrices belonging to $d_1 \otimes d_2$ of rank $k \in \{2, 3, \dots, d_1 d_2\}$. We study the variation in the ability of a chosen criterion to detect entanglement with respect to the two parameters :


\begin{enumerate}
    \item \textbf{Dependence on rank:} For a fixed total system of dimension \(d_1 d_2\), how does the performance of the different entanglement criteria vary with the rank, \(k\), of the states?
    
    \item \textbf{Effect of subsystem dimensions:} What effect do the dimensions of the individual subsystems have on the efficacy of a criterion for a given rank?
\end{enumerate}

\textit{Objectives.} We analyze the performance of individual criteria, with respect to variation in the aforementioned parameters, and provide a comparative assessment among them. The {\it performance} of a criterion refers to whether or not it can successfully detect an NPT entangled state. Its efficacy is gauged through two principal factors: $(1)$ the fraction of randomly generated NPT states that the criterion can conclude as entangled, i.e., the ratio of the number of states identified as entangled by the given criterion to the total number of sampled NPT states; and $(2)$ the entanglement content of the state, investigated through the mean and minimum value of the logarithmic negativity~\cite{Vedral1997_prl, Plenio_PRL_2005_log-neg} (as defined in Sec.~\ref{sec:bounds_ent_success}), over all simulated NPT states. A high fraction of successful detections and a low entanglement content for the same means that the criterion is {\it effective}. Furthermore, if a criterion admits a higher fraction of successful detection and/or a lower necessary entanglement content than another, the former is said to be {\it relatively more effective}.

To quantify entanglement, we define the normalized logarithmic negativity as
$
\mathrm{LN}(\rho_{12}) = 
\frac{\log_2 \left\| \rho_{12}^{T_1} \right\|_1}
{\log_2 \big( \min(d_1, d_2) \big)},
$
where $\|A\|_1=\trace{\sqrt{A^\dagger A}}$ represents the trace norm. The normalization factor is so chosen because LN is upper bounded by the determinant of the density matrix of the smaller subsystem, which is further upper bounded by the logarithm of its dimension, i.e., $\mathrm{LN} \leq \det (\rho_{\min(d_1, d_2)}) \leq \min(d_1, d_2)^{-\min(d_1, d_2)} \leq \log_2 (\min(d_1, d_2))$. This ensures $0\leq\mathrm{LN}(\rho_{12})\leq1.$

\subsection{Connection between rank and dimension in entanglement detection}
\label{sec:sec_IIIA}

Before answering the questions for randomly generated states, we first provide bounds on the rank of bipartite mixed states in arbitrary dimensions. This determines whether their entanglement can be successfully detected by using two of the considered measures - the entropy and the realignment criteria. Specifically, we establish how the entropy criterion is grossly unsuitable for evaluating whether a high-rank state is entangled or separable, while the realignment criterion fares better when the dimensions of the subsystems grow. 

{\it The entropy criterion is weak in detecting high-rank states.} 
Given its operational significance as conditional entropy, whose negativity has been proven to be a resource in several quantum information processing tasks~\cite{Bennett_1992_prl, Devetak_PRSA_2005_secret-key, Horodecki_CMP_2006_negative-conditional-entropy-state-merging, Berta_NP_2010_negative-conditional-entropy-quantum-memory, Yang_PRL_2019_negative-conditional-entropy-randomness-distillation}, it is worthwhile to analyze when the entropy criterion is successful in predicting the entanglement of arbitrary finite-dimensional random bipartite states on average. Our following proposition attempts to quantify its efficacy.

\begin{proposition}
\label{pro:entropy_full_rank}
{\it For any bipartite system in \(d_1 \otimes d_2\), the entropic criterion, \(\mathrm{E}\), fails to detect the entanglement of Haar uniformly generated states, when the rank, $k > \max(d_1, d_2)$.}  
\end{proposition}

\textit{Proof.} 
Without loss of generality, consider $d_2 \geq d_1$. Recall also that the entropy criterion concludes that a state, $\rho_{12}$, acting on $d_1 \otimes d_2$, is separable if both $S_{12} - S_1 \geq 0$ and $S_{12} - S_2 \geq 0$ and hence the violation implies entanglement in a given state. 
For the average entropies (where the averaging is performed over the Haar measure), it is known that~\cite{Page_PRL_1993_average-entropy, Sen_PRL_1996_average-entropy-proof, Foong_PRL_1994_average-entropy-proof, Kendon2002_pra}
\begin{eqnarray}
    \langle S_1 \rangle &\approx& \log d_1 - \frac{d_1}{2 d_2 k}, \,\,\, 
    \langle S_2 \rangle \approx \log d_2 - \frac{d_2}{2 d_1 k}, and \nonumber \\
    \langle S_{12} \rangle &\approx& \log k - \frac{k}{2 d_1 d_2} \label{eq:entropy_proof_expression},
\end{eqnarray}
with \(k\) being the rank of the density matrix, \(\rho_{12}\).
Let us first concentrate on \(\langle S_{12}\rangle-\langle S_{2}\rangle\). It vanishes at $k = d_2$, and in general, its derivative reads
\begin{eqnarray}
    \frac{d}{dk} (\langle S_{12}\rangle - \langle S_2\rangle)= \frac{2d_{1}d_{2}k-k^{2}-d_{2}^{2}}{2d_{1}d_{2}k^{2}}.
    \label{eq:derivative_entropy}
\end{eqnarray}
Note that the denominator in Eq.~\eqref{eq:derivative_entropy} is always positive. Further, the numerator is also positive definite for $k \in \{2, 3, \dots, d_1 d_2\}$ and its derivative satisfies, $2(d_1 d_2 - k) \geq 0 ~\forall~ k \leq d_1 d_2$. Therefore, $\langle S_{12}\rangle - \langle S_2\rangle$ has a strictly increasing derivative for all ranks $k \in \{d_2, d_2 + 1, \dots, d_1 d_2\}$, which implies that it is monotonically increasing. Therefore, with $(\langle S_{12}\rangle-\langle S_{2}\rangle)|_{k = d_2} = 0$, we have $\langle S_{12}\rangle - \langle S_2\rangle \geq 0$ when $k \in \{d_2, d_2 + 1, \dots, d_1 d_2\}$. This completes the first part of our proof.

Our job now is to prove that $\langle S_1\rangle \leq \langle S_2\rangle$ when $d_2 \leq k \leq d_1 d_2$. This would allow us to show that $\langle S_{12}\rangle - \langle S_2\rangle \geq 0 \implies \langle S_{12}\rangle - \langle S_1\rangle \geq 0$, and thus the entropy criterion fails to conclude the state to be entangled. Let $d_2 = d_1 + x$ with $x \in [0, d_1 d_2 - d_2]$. We note that 
\begin{eqnarray}
   \nonumber (\langle S_2\rangle - \langle S_1\rangle)|_{k = d_1 + x} = \log \frac{d_1 + x}{d_1} - \frac{1}{2 d_1} + \frac{d_1}{2(d_1 + x)^2},\\
    \label{eq:S21_diff}
\end{eqnarray}
and $(\langle S_2\rangle - \langle S_1\rangle)|_{k = d_2 = d_1} = 0$ by substituting $x = 0$ in Eq.~\eqref{eq:S21_diff}. Moreover, the derivative with respect to $x$, $\Big((d_1 + x)^2 - d_1 \Big)/\Big( d_1 + x \Big)^3 \geq 0 ~\forall~ x$ and thus $\langle S_2 \rangle - \langle S_1 \rangle$ is a monotonically increasing function of $x$, and is hence always positive semidefinite for all $k \in (d_2, d_1 d_2)$. Taking its derivative, this time with respect to \(k\), it is evident that
\begin{equation}
    \frac{d}{dk}(\langle S_{2}\rangle-\langle S_{1}\rangle) = \frac{d_{2}^{2}-d_{1}^{2}}{2d_{1}d_{2}k^{2}}\geq 0 \quad \forall d_{2}\geq d_{1}.
    \label{eq:S21_derivative_k}
\end{equation}
Thus \((\langle S_{2}\rangle-\langle S_{1}\rangle)\geq 0\) at \(k=d_{2}\) and is monotonically increasing, which suggests \( \langle S_1\rangle \leq \langle S_2\rangle ~\forall~ k \in \{d_{2}, d_2 + 1, \dots, d_{1}d_{2}\} \). This completes the proof. 
$\hfill \blacksquare$

\noindent {\bf Remark $\mathbf{1}$.} The expressions for $\langle S_{1} \rangle$,$\langle S_{2} \rangle$ and $\langle S_{12} \rangle$, i.e. the average von Neumann entropies of $d_1$ dimensional subsystem, $d_2$ dimensional subsystem and the whole $d_1d_2$ dimensional system respectively, are approximated for $d_1d_2\gg0.$ 
We observe from our simulations that the fraction of states detected as entangled by the entropy criterion is merely $0.055$ for $d_1 = 2$ and vanishes beyond $d_1 = 4$ when we fix $d_2 = 8$ and $k = 9$.

The aforementioned proposition establishes that the entropy criterion is very weak in terms of detecting entanglement of mixed states, especially if both $d_1, d_2 \gg 1 $, where it fails to detect states in a wide range of rank $k \in (d_2, d_1 d_2)$. This can again be confirmed from the numerical simulations performed for random states in the succeeding sections.

\textit{The efficacy of the realignment criterion.} Apart from the realignment criterion's capability of detecting PPT-bound entangled states, it also provides necessary criteria to characterize entanglement-breaking channels~\cite{Lupo_JPA_2008_realignment-schmidt}. Compared to the entropy criterion, it manages to identify a relatively higher fraction of states, as will be verified later with the numerical simulations. In the following, we establish exact bounds on the rank of a state such that realignment is effective in concluding its separability.
\begin{proposition}
\label{pro:real_detec}
{\it The realignment criterion, \(\mathrm{R}_L\), becomes ineffective for detecting entanglement in a bipartite state, \(\rho_{12}\) acting on \(d_1 \otimes d_2\), when its rank satisfies \(k \geq \frac{d_{1}^{3} d_{2} - 1}{d_{1}(d_{2} - d_{1})}.\)}
\end{proposition}
\noindent {\it Proof.}
To check the realignment criterion, we need to consider the realignment matrix, \(\rho^{\mathrm{R}_L}_{12}\), of the given density matrix, $\rho_{12}$. The state \(\rho_{12}\) is separable when~\cite{chen2003matrixrealignment, realighment_rudo_2003}
  \( \sum_i g_i = \|\rho^{\mathrm{R}_L}_{12}\|_1 \leq 1\), 
where $g_i$s are the singular values of the realignment matrix, and the violation of the bound
indicates entanglement. Since the matrix $\rho^{\mathrm{R}_L}_{12}$ has dimension $d_1^2 d_2^2$, its rank, $r^{\mathrm{R}_L} \leq \min(d_1^2, d_2^2) = d_1^2 $, assuming \(d_2 \geq d_1\) without loss of generality. Using the relation between the Frobenius norm, $\left\| \cdot \right\|_F$~\cite{nielsen_2010, Watrous_2018}, and the trace norm, we can write \(\left\| \rho^{\mathrm{R}_L}_{12} \right\|_1 \leq d_1 \left\| \rho^{\mathrm{R}_L}_{12} \right\|_F,
\) where, $\left\| \cdot \right\|_F = \sqrt{ \sum_{i,j} |a_{ij}|^2 }$ for a matrix with elements $a_{ij}$. Since $\rho^{\mathrm{R}_L}_{12}$ has the same elements as $\rho_{12}$, we obtain
\begin{eqnarray}
\left\| \rho^{\mathrm{R}_L}_{12} \right\|_1 \leq d_1 \left\| \rho_{12} \right\|_F = d_1 \sqrt{ \text{Tr}(\rho_{12}^2) },
\label{eq:Realign_2}
\end{eqnarray}
where the last equality follows from $\sum_{i,j} |a_{ij}|^2 = \Tr (\rho_{12}^\dagger \rho_{12}) = \Tr \rho_{12}^2$, which is nothing but the state's purity, given as \(\operatorname{Tr}(\rho_{12}^2) = \frac{d_1 d_2 + k}{d_1 d_2 k + 1},\)~\cite{Kendon2002_pra} for \(\rho_{12}\) on \(d_1 \otimes d_2\) of rank $k$. This leads to
\begin{eqnarray}
 \left\| \rho_{12}^{\mathrm{R}_L} \right\|_1 \leq d_1 \sqrt{ \frac{d_1 d_2 + k}{d_1 d_2 k + 1} }= \sqrt{f_{\mathrm{R}_L}(k, d_1, d_2)},
\label{eq:Realign_4}
\end{eqnarray}
with \(
f_{\mathrm{R}_L}(k, d_1, d_2) = d_1^2 \left( \frac{d_1 d_2 + k}{d_1 d_2 k + 1} \right) > 0, ~\forall~ k \in (2, d_1 d_2)
\). Taking its derivative, we find that \( \frac{d f_{\mathrm{R}_L}}{d k} = d_1^2 (1 - d_1^2 d_2^2)/(1 + d_1 d_2 k)^2 < 0, ~\forall~ d_1, d_2 \geq 2 ~\text{and}~ k \in (2, d_1 d_2)\). Since the derivative of a function and its square-root have the same sign in regions where it is positive definite, \(\sqrt{f_{\mathrm{R}_L}(k, d_1, d_2)}\) is a monotonically decreasing function of \(k\). Let \(\sqrt{f_{\mathrm{R}_L}(k = k_0, d_1, d_2)} = 1\) whence its monotonicity implies \(\sqrt{f_{\mathrm{R}_L}(k \geq k_0, d_1, d_2)} \leq 1\). It is easy to derive \(k_0 = \frac{d_1^3 d_2 - 1}{d_1 (d_2 - d_1)} \) and the proof follows. $\hfill \blacksquare$

\aka{
\subsection{Equivalence  of $\mathcal{P}$ and $\mathrm{R}_D$ criterion in \(2\otimes d\)}

Before proceeding to our numerical results, let us focus on two more entanglement detection criteria, viz, $\mathcal{P}$ and $\mathrm{R}_D$. While it has already been established that the two criteria are equivalent for $2 \otimes d$ systems~\cite{JBatle_2004_iop_allconnnection}, in this subsection, we provide an independent proof of the same.

\begin{proposition}
     {\it \(\mathcal{P}\) and \(\mathrm{R}_D\) are equivalent in \(2\otimes d_2\).}
     \label{pro:ppt_red_eqv}
\end{proposition}

\noindent {\it Proof.} Let \(\rho_{12}\) be a bipartite density matrix acting on \({d_1} \otimes {d_2}\), which, in the computational basis, using the spectral decomposition~\cite{nielsen_2010, Preskill, wilde_2013, Watrous_2018}, can be represented as
\begin{eqnarray}
&&\rho_{12} = \sum_{i,j = 0}^{d_1 - 1} \sum_{\mu,\nu = 0}^{d_2 - 1} a_{ij\mu\nu} \ket{i\mu}\bra{j\nu} 
= \sum_{\mu,\nu=0}^{d_2 - 1} \big( a_{00\mu\nu} \ket{0\mu}\bra{0\nu} \nonumber \\ 
&& + a_{01\mu\nu} \ket{0\mu}\bra{1\nu} 
 +\; a_{10\mu\nu} \ket{1\mu}\bra{0\nu} + a_{11\mu\nu} \ket{1\mu}\bra{1\nu} \big),\nonumber\\
\label{eq:bipartite_state_2_d}
\end{eqnarray}
where, in the second equation, we set $d_1 = 2$. 
The partial transpose, \(\rho_{12}^{T_1}\), with respect to subsystem \(1\) is given by
\(\rho^{T_1}_{12} = \sum_{\mu,\nu=0}^{d_2 - 1} \big( a_{00\mu\nu} \ket{0\mu}\bra{0\nu} + a_{10\mu\nu} \ket{0\mu}\bra{1\nu}
 +\; a_{01\mu\nu} \ket{1\mu}\bra{0\nu} + a_{11\mu\nu} \ket{1\mu}\bra{1\nu} \big)\).
A necessary condition for separability of \(\rho_{12}\) is $\rho^{T_1}_{12} \geq 0$.
On the other hand, according to \(\mathrm{R}_D\), the separability conditions read
\(\rho_1 \otimes I_2 - \rho_{12} \geq 0\) and \(I_1 \otimes \rho_2 - \rho_{12} \geq 0\). Let us consider the operator $r = I_1 \otimes \rho_2 - \rho_{12}$ in $2 \otimes d_2$, written as
\begin{eqnarray}
r &=& \sum_{\mu,\nu=0}^{d_2 - 1} \big( a_{00\mu\nu} \ket{1\mu}\bra{1\nu} - a_{01\mu\nu} \ket{0\mu}\bra{1\nu} \nonumber \\
&& -\; a_{10\mu\nu} \ket{1\mu}\bra{0\nu} + a_{11\mu\nu} \ket{0\mu}\bra{0\nu} \big).
\label{eq:M_2_d}
\end{eqnarray}
We note that \(r\) and \(\rho_{12}^{T_1}\) are related by a unitary transformation, \( r = U^\dagger \rho^{T_1}_{12} U\), where \(U=I_1\otimes \sigma_2^y\) with \(\sigma^y\) being the Pauli-y matrix. Thus, they have the same eigenvalues. Consequently, if one of them is positive semidefinite, so is the other, while if one has at least one negative eigenvalue, the other also exhibits the same property. From the above argument, we can conclude that in \(2 \otimes d_2\), \(\mathcal{P}\) and \(\mathrm{R}_D\) are equivalent. $\hfill \blacksquare$
}

\subsection{Capability of entanglement criteria for typical states}

We now move on to numerically evaluating and comparing the ability of the various criteria in identifying entanglement across a large number of Haar uniformly generated states.

\subsubsection{Benchmarking  via partial transposition}

The comparison between the entanglement detection methods will be carried out with respect to partial transposition $\mathcal{P}$. We focus exclusively on states that are entangled under the partial transposition criterion, i.e., only non-positive partial transpose (NPT) states are considered. An immediate question arises regarding this choice since $\mathcal{P}$ is not a necessary condition for arbitrary finite-dimensional bipartite systems~\cite{HORODECKI19961}. One main reason is that, for all practical purposes, such as information-theoretic protocols~\cite{Bennett_1992_prl, Bennett_1993_prl, Gisin_2002_RMP}, quantum computation~\cite{Jozsa_arXiv_1997_entanglement-quantum-computation}, and even physical realization through many-body systems
~\cite{Srivastava_book_2024_many-body-entanglement}, distillable NPT states~\footnote{Note that NPT bound entangled states have been conjectured and some evidence for them also exists~\cite{DiVincenzo_PRA_2000_NPT-bound, Dur_PRA_2000_NPT-undistillable}} are found to be useful.
It is therefore natural to focus on NPT entangled states. While PPT-bound entangled states have been shown to offer operational benefits in secret key generation~\cite{Horodecki_PRL_2005_bound-entanglement-key}, steering~\cite{Moroder_PRL_2014_bound-entanglement-steering}, and metrology~\cite{Pal_PRR_2021_bound-entanglement-metrology}, none of the criteria considered in our work, except realignment, are capable of detecting them. Therefore, incorporating the detection ability of bound entangled states would not allow us to compare the criteria on a balanced footing. In particular, it would create a clear distinction between realignment and the other criteria, since a trivial superiority of the realignment criterion would emerge from its ability to detect such states, which we will show to be not true for NPT states. Thus, we constrain ourselves to NPT entangled states in this work, and using the Peres-Horodecki separability criterion~\cite{peres_prl_1996_ppt, HORODECKI19961} as a benchmark and the logarithmic negativity for quantification, we establish a hierarchy among different qualitative entanglement detection criteria. As a second justification of our choice, let us argue how an overwhelming majority of states in $d_1 \otimes d_2$, of rank $k$, exhibit non-positivity under partial transposition, when obtained from a global pure state in $d_1 \otimes d_2 \otimes k$ through partial tracing. The probability that a two-qudit mixed state is NPT is extremely high if $N \leq 3m - 2$~\cite{Kendon2002_pra}, where $N$ is the size of the global pure state, i.e., $N = 3$ in our case, and $m = 2$ represents the number of qudits in the state under consideration. Thus, the condition $N \leq 4$ is always satisfied for the simulated states in our work, thereby ensuring a high probability of NPT states. Moreover, a sufficient condition for such a state to have positive partial transpose (PPT) is $k \geq d_1 d_2 (d_1 d_2 - 1) - 1 \approx d_1^2 d_2^2$~\cite{Kendon2002_pra}, which is never satisfied for large $d_1 d_2$ since $2 \leq k \leq d_1 d_2$. Let us now briefly discuss how the random density matrices, \(\rho_{12}\), are generated in \(d_1\otimes d_2\).


\subsubsection{ Generation of bipartite mixed states}

We simulate \textcolor{black}{\( 10^{6}\)} random states, sampled uniformly according to the Haar measure~\cite{bengtsson_zyczkowski_2006}. Each mixed state is obtained by tracing out the third subsystem from a random pure state, \(|\psi\rangle_{123} = \sum_{l,m,k} (a_{lmk} + i b_{lmk}) \, |l\rangle_1 \otimes |m\rangle_2 \otimes |k\rangle_3 \), defined on \({d_1} \otimes {d_2} \otimes {k}\). The coefficients, \(a_{lmk}\), and, \(b_{lmk}\), are sampled independently from a Gaussian distribution, \(\mathcal{G}(0,1)\), with vanishing mean and unit standard deviation. The resulting reduced state, \(\rho_{12} = \mathrm{Tr}_{3}[(|\psi\rangle \langle \psi|)_{123}]\), is a bipartite mixed state of rank, \(k\)~\cite{Ozols_essay_2009, Zyczkowski_JMP_2011, Dahlsten_JPA_2014} (see also~\cite{Banerjee_PRA_2020_decoherence-localisable-entanglement}). Apart from $2 \otimes 2$ and $2 \otimes 3$ dimensions, among the generated states, a fraction of states are also PPT \cite{zyczkowski_pra_1998}, which we discard and consider only NPT states for calculating other entanglement detection methods. Although all the results presented hereafter are for $5\times10^5$ states, we also numerically simulate $10^4,5\times 10^4,10^5$ and $5\times 10^5$ Haar random states and check whether all the results reported converge up to 3 decimal places. Indeed, we find that the results obtained with $10^5 $ and $5\times10^5$ converge up to $3$ decimal places (see Sec. \ref{app:convergence})

\subsubsection{Hierarchy among entanglement identification conditions}

To establish the hierarchy among criteria belonging to \(\mathcal{S}\), we now introduce the notion of fraction of successfully detected randomly generated states with respect to the nonvanishing \(\text{LN}\).  
For a quantitative comparison, we define a normalized fraction of a given entanglement criterion, \(\mathcal{E}\), as 
\begin{eqnarray}
F^\mathcal{E}
= \frac{\sum_{i} i_\mathcal{E}}{\sum_{i} i \Big|_{\mathrm{LN}(i)>0}}.
\label{eq:frac_state_det}
\end{eqnarray}
where we denote the states successfully detected by a criterion $\mathcal{E}$ by $i_\mathcal{E}$, and the randomly generated states which are NPT entangled, i.e., have non-zero logarithmic negativity (LN), are denoted as $i \Big|_{\mathrm{LN}(i) > 0}$. By definition, \(0\leq F^\mathcal{E} \leq 1\). Trivially, for the partial transposition rule, \(F^{\mathcal{P}}=1\) and the equivalence of reduction  and \(\mathcal{P}\) criteria in \(2\otimes d\) \cite{chen2003matrixrealignment, JBatle_2004_iop_allconnnection} implies \(F^{\mathrm{R}_D}=1\) too. Given a dimension and rank, if $F^{\mathcal{E}_1} > F^{\mathcal{E}_2}$, we can surely infer that the criterion $\mathcal{E}_1$ is more effective in identifying typical entangled states than $\mathcal{E}_2$.

\begin{table}[H]
\centering
\renewcommand{\arraystretch}{1.2} 
\begin{tabular}{|>{\centering\arraybackslash}p{2cm}|
                >{\centering\arraybackslash}p{2cm}|
                >{\centering\arraybackslash}p{2cm}|
                >{\centering\arraybackslash}p{2cm}|}
\hline
\textbf{\(k\)} & \textbf{$F^\mathcal{M}$} & \textbf{$F^\mathrm{E}$} & \textbf{\(F^{\mathrm{R}_L}\)} \\
\hhline{|====|}
$2$  & $1$     & $0.999$   & $1$      \\ \hline
$3$  & $1$     & $0.998$   & $0.999$  \\ \hline
$4$  & $1$     & $0.946$   & $0.968$  \\ \hline
$5$  & $0.799$ & $0.494$   & $0.563$  \\ \hline
$6$  & $0.510$ & $0.054$  & $0.145$  \\ \hline
$7$  & $0.333$ & $0.001$   & $0.021$  \\ \hline
$8$  & $0.223$ & \textemdash & $0.002$  \\ \hline
$9$  & $0.150$ & \textemdash & \textemdash \\ \hline
$10$ & $0.099$ & \textemdash & \textemdash \\ \hline
\end{tabular}

\captionsetup{justification=Justified,singlelinecheck=false}
\caption{Normalized fraction, \(F^{\mathcal{E}}\), of entangled states detected by a criterion, \(\mathcal{E}\), belonging to the set \(\mathcal{S}\) in Eq.~(\ref{eq:measure_set}). Here we consider a \(2 \otimes 5\) dimensional system, with rank $2 \leq k \leq 10$. A dash (\textemdash) indicates that the respective criterion fails to detect any entangled state. Since \(F^{\mathcal{P}} = F^{\mathrm{R}_D}=1\), we skip them. \aka{It is evident that majorization performs better than the realignment for qubit-qudit systems for all ranks, while entropy and realignment fail to identify entangled states when the rank of the states is high, as proven in Propositions \ref{pro:entropy_full_rank} and \ref{pro:real_detec}.}}
\label{tab:frac_state_detection}
\end{table}

The results obtained through numerical simulations are presented in two sectors -- \((1)\) random states in \(2\otimes d\) with \(d>3\) and \((2)\) states in \(d_1 \otimes d_2\) with \(d_1,d_2\geq3\) \(\forall k\). The reason for such division is to better understand the behavior of the entanglement detection method in \(2\otimes d\) as compared to the states in \(3\otimes d\) onwards. For example, the question of the existence of bipartite NPT-bound entangled states is still unresolved above qubit-qudit systems~\cite{DiVincenzo_PRA_2000_NPT-bound, Dur_PRA_2000_NPT-undistillable}. We will now show that our study can reveal some unknown characteristics of the set of criteria \(\mathcal{S}_1=\{\mathcal{M}, E, \mathrm{R}_L\}\) for random states.

\begin{figure}[h]
    \centering \includegraphics[width=\linewidth]{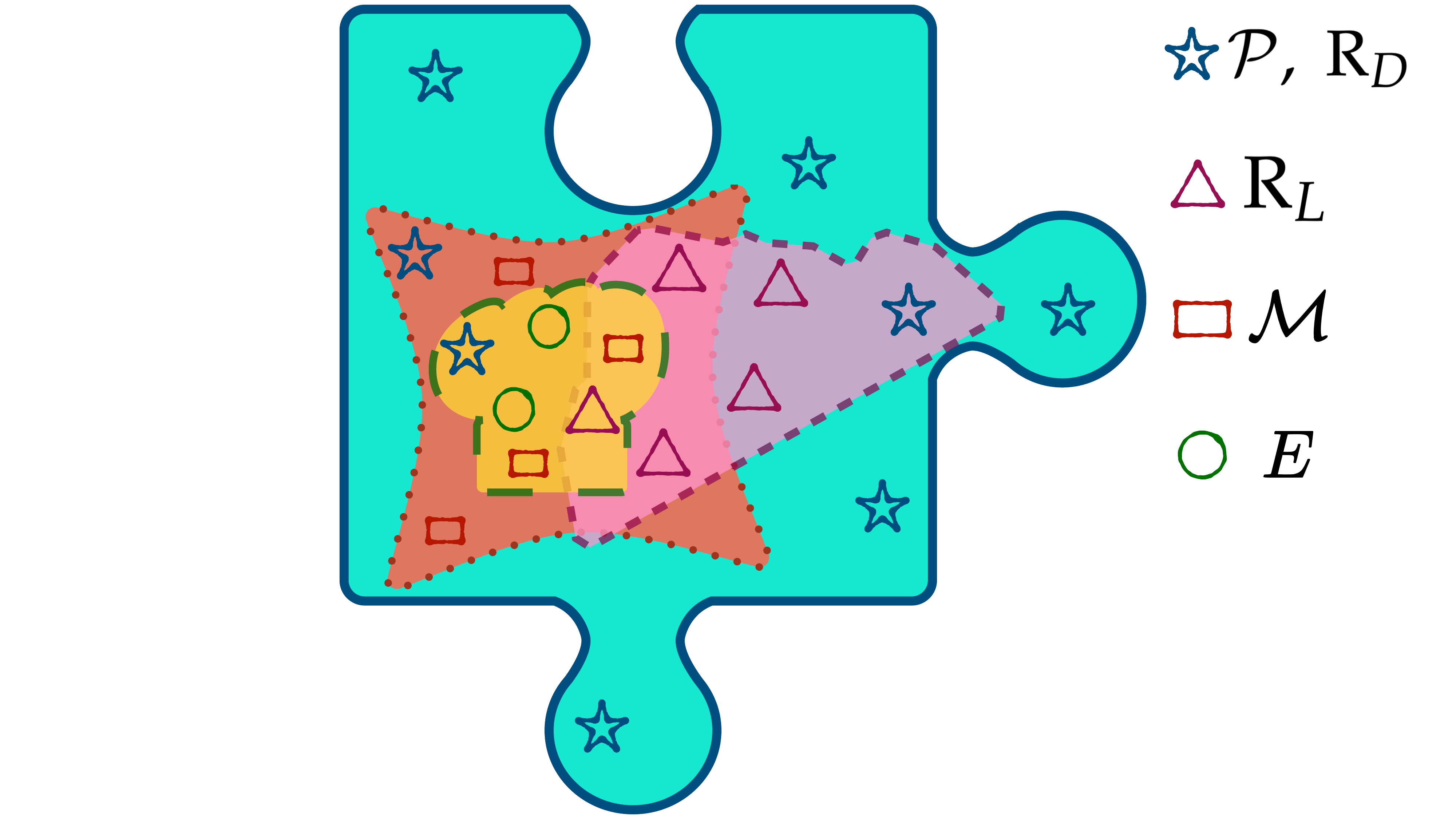}
    \captionsetup{justification=Justified,singlelinecheck=false}  \caption{ Pictorial hierarchy of entanglement detection criteria for random rank-$6$ states in a $2\otimes 6$ bipartite system. Colored regions indicate sets of states detected as entangled by different criteria: PPT \(\mathcal{P}\) and Reduction criteria, \(\mathrm{R}_D\) (blue with solid boundary), Majorization, \(\mathcal{M}\) (red with dotted boundary), the entropy-based criterion, \(E\) (yellow with dashed boundary), and the realignment criterion, \(\mathrm{R}_L\) (purple with dash-dotted boundary). Each point in the figure represents a randomly generated density matrix which is detected as entangled by $\mathcal{P}$ and $\mathrm{R}_D$ (stars), $\mathrm{R}_L$ (triangles), $\mathcal{M}$ (rectangles), and $E$ (circles). The relative sizes and the nested structure of the colored regions illustrate a clear hierarchy in \emph{entanglement detection power}, with $\mathcal{P}$ and $\mathrm{R}_D$ detecting the largest fraction of entangled states, followed by $\mathcal{M}$, then $\mathrm{R}_L$, and finally $E$. The partial overlap between the colored regions indicates that the same entangled state can be detected by multiple criteria simultaneously. These overlaps reflect common detection capability rather than any strict set-theoretic inclusion among the criteria, and the hierarchy is therefore understood purely in terms of comparative entanglement detection power.
}
\label{fig:criteria_diag}
\end{figure}

{\it Ability to determine entanglement in \(2\otimes d\).} We observe and compare some distinct features for \(\mathcal{M}, E\) and \(\mathrm{R}_{L}\). As expected, \(E\) is the weakest criterion among all. In particular,  we find that when 
\(k \gtrsim d_2+2\) (\(d_2 \geq d_1=2\)), \aka{almost no randomly generated states can be detected by the entropic criteria (see Table \ref{tab:frac_state_detection} for \(2\otimes 5\) states), an observation which also vindicates Proposition \ref{pro:entropy_full_rank}}. 
On the other hand, it was shown that both separable and PPT-bound entangled states satisfy the majorization criterion, while the realignment condition can detect PPT-bound entangled states. Therefore, one can be tempted to infer that majorization will be weaker than realignment for the identification of entangled states. We show that this is not true in the qubit-qudit case. Specifically, we find that although the fraction of NPT states detectable by \(\mathcal{M}\) and \(R_L\) decreases with increasing rank for a fixed \(2\otimes d\) system, majorization is more effective in determining entangled states than realignment, as shown in Table~\ref{tab:frac_state_detection}. Precisely, for high-rank states, the realignment criterion fails to detect any distillable states and \(F^{\mathrm{R}_L}\to 0\) with \( k \sim 2 d\) . This allows us to establish a clear hierarchy of the effectiveness in entanglement detection methods, from most to least powerful in \(2 \otimes d\) typical systems, as
\begin{eqnarray}
    F^\mathcal{P} \equiv F^{\mathrm{R}_D} > F^\mathcal{M} > F^{\mathrm{R}_L} > F^\mathrm{E},
    \label{eq:hierarchy_1}
\end{eqnarray}
where the symbol ``\(>\)'' denotes a greater entanglement detection power.  Hence, the set of states detectable by different criteria follows the ordering \(\mathcal{P}=\mathrm{R}_D\supset \mathcal{M}\supset \mathrm{R}_L \supset E\) for generic states, where the notations for the criteria are also used to denote the corresponding set of identifiable states. Note that, low-rank states can be detected by all the criteria, whereas entangled states with very high rank (close to the total dimension, $2 \times d$) are only successfully detected by $\mathcal{P}$ and $\mathrm{R}_D$. To have an intuitive representation of the detection power of the different criteria, we present a pictorial illustration for states of intermediate rank, $(2\times d/2)$, in $2 \otimes d$ in Fig.~\ref{fig:criteria_diag}. It is noticeable that while some states are detected by all the considered criteria, others are identified only by some. Importantly, this ordering reflects a typical inclusion of detectable states rather than a strict set-theoretic containment for every individual state.

{\it Ordering among entanglement detection criteria in \(d_1 \otimes d_2\) with \(d_1,d_2\geq 3\).} The common observation between states in \(d_1\otimes d_2\) with \((d_1>2\) \text{and} \(d_2>3)\) and \(2\otimes d_2\) is that \(F^{\mathcal{E}}\) decreases with the rank for a fixed total dimension.
The differences between two qudits and qubit-qudit systems are as follows:
$(1)$ The PT and the reduction criteria are no longer equivalent, and the performance of the latter gets worse with the increase of the rank.  $(2)$ For a fixed \(d_1\), with increasing \(d_2\), all the criteria become equally effective and in this case, \(F^{\mathrm{R}_L}\geq F^{\mathrm{R}_D}\geq F^{\mathcal{M}}\geq F^\mathrm{E}\) for low \(k\) (see Table \ref{tab:sub_dim_and_rank_vary}). $(3)$ The significant contrasting behavior emerges for moderate rank states -- realignment performs better than majorization which is not true in \(2 \otimes d\). Surprisingly, we observe that even for a fixed \(d_1\geq 3\), there exists \(k\) for which \(F^{\mathrm{R}_L}\geq F^{\mathrm{R}_D}\), thereby confirming a stronger performance by the realignment criterion compared to the others in \(\mathcal{S} \backslash  \mathcal{P}\) (see Table \ref{tab:sub_dim_and_rank_vary}).  We will also observe in the next subsection that the local dimension can also be responsible for the hierarchy among criteria, apart from dimension and rank. 
$(4)$ For full rank states, only reduction criteria can identify some NPT entangled states while entropy, majorization, and realignment fail, i.e., \(F^{\mathrm{R}_D} > F^{\mathrm{R}_L} > F^{\mathcal{M}}> F^{E}\).

\begin{table}[H]
\centering
\renewcommand{\arraystretch}{1.3}
\begin{tabular}{|>{\centering\arraybackslash}p{1.3cm}|>{\centering\arraybackslash}p{1.3cm}|>{\centering\arraybackslash}p{1.3cm}|>{\centering\arraybackslash}p{1.3cm}|>{\centering\arraybackslash}p{1.3cm}|>{\centering\arraybackslash}p{1.3cm}|}
\hline
\textbf{\(d_1 \otimes d_2\)} & \textbf{\(k\)} & \textbf{$F^\mathcal{M}$} & \textbf{$F^\mathrm{E}$} & \textbf{$F^{\mathrm{R}_L}$} & \textbf{$F^{\mathrm{R}_D}$} \\
\hhline{|======|}

$3 \otimes 3$ & $9$ & $0.004$ & \textemdash & $0.743$ & $0.352$ \\ \hline

\multirow{2}{*}{$3 \otimes 4$} 
 & $9$  & $0.014$  & \textemdash & $0.968$ & $0.902$ \\ \cline{2-6}
 & $12$ & $0.001$  & \textemdash & $0.330$ & $0.424$ \\ \hline

\multirow{2}{*}{$3 \otimes 5$} 
 & $9$  & $0.049$  & \textemdash & $0.995$ & $0.999$ \\ \cline{2-6}
 & $15$ & \textemdash & \textemdash & $0.005$ & $0.423$ \\ \hline

\multirow{2}{*}{$3 \otimes 6$} 
 & $9$  & $0.124$  & \textemdash & $1$ & $1$ \\ \cline{2-6}
 & $18$ & \textemdash & \textemdash & \textemdash & $0.701$ \\ \hline

\end{tabular}

\captionsetup{justification=Justified,singlelinecheck=false}
\caption{Normalized fraction, \(F^\mathcal{E}\), of random NPT entangled states for various detection criteria \(\mathcal{E}\) applied to states in \(3 \otimes d_2\) systems, with two different ranks,  \(k = d_1d_2\) and \(k > d_1d_2/2\) (moderate rank).   All other specifications are the same as in Table~\ref{tab:frac_state_detection}. For moderate rank, realignment clearly performs better than reduction. This is a contrasting behavior compared to the qubit-qudit systems. }
\label{tab:sub_dim_and_rank_vary}
\end{table}

\section{Bounds on entanglement necessary for successful detection}
\label{sec:bounds_ent_success}

While the normalized fraction of detected entanglement, $F^\mathcal{E}$, sheds light on whether the criterion could successfully detect entanglement or not, we are now interested in addressing the question -- ``{\it does the efficiency of the entanglement detection method depend on the amount of entanglement possessed by random quantum states?}'' We answer it affirmatively by introducing two figures of merit, the mean entanglement, \(M_{\mathcal{E}}\) and the minimum entanglement, \(m_{\mathcal{E}}\), where the entanglement is quantified by \(\text{LN}\). In particular, they are defined for a given criterion, \(\mathcal{E}\), as 
\begin{eqnarray}
 M_{\mathcal{E}} = \frac{1}{n} \sum_{\substack{i = 1 \\ \mathcal{E}(i)}}^{n} \mathrm{LN}(i),
    \text{and} \quad
     m_{\mathcal{E}} = \min_{\substack{i = 1 \\ \mathcal{E}(i)}}^n [\mathrm{LN}(i)], 
    \label{eq:min_ent} 
\end{eqnarray}
where the summation and minimum are performed only when the entanglement detection method can identify the state, \(i\), as entangled, and we calculate the entanglement content solely for these states with \(n\) being the total number of states for which the criteria, \(\mathcal{E}\), is successful. The mean and minimum detectable entanglement can provide bounds on the resource content of the state necessary for successful entanglement detection. The smaller the minimum entanglement, $m_\mathcal{E}$, the lower is the entanglement that can be detected by the criterion, $\mathcal{E}$, and, therefore, the more adept it is. There is a subtlety to note, however. Since the entanglement content decreases with the rank, $k$~\cite{Gupta_PRA_2022_non-Markovianity, Gupta_PRA_2022_catalysis}, a lower minimum entanglement at lower ranks proves that the criterion is more powerful than one with a higher minimum entanglement. On the other hand, the mean entanglement not only depends on the entanglement content of the states detected, but also on the number of states that are detected, i.e., on $F^\mathcal{E}$. A higher mean entanglement can be due to two factors - greater entanglement due to lower rank and higher total dimensions~\cite{Gupta_PRA_2022_catalysis}, and a lower fraction of detected states. In this case, a higher mean entanglement at higher ranks necessarily means that a smaller number of states are detected, and thus the criterion is less effective.

\subsection{Detection strength of various measures through mean and minimum entanglement}
\label{subsec:det_mean_min}

\begin{figure}[h]
    \centering \includegraphics[width=\linewidth]{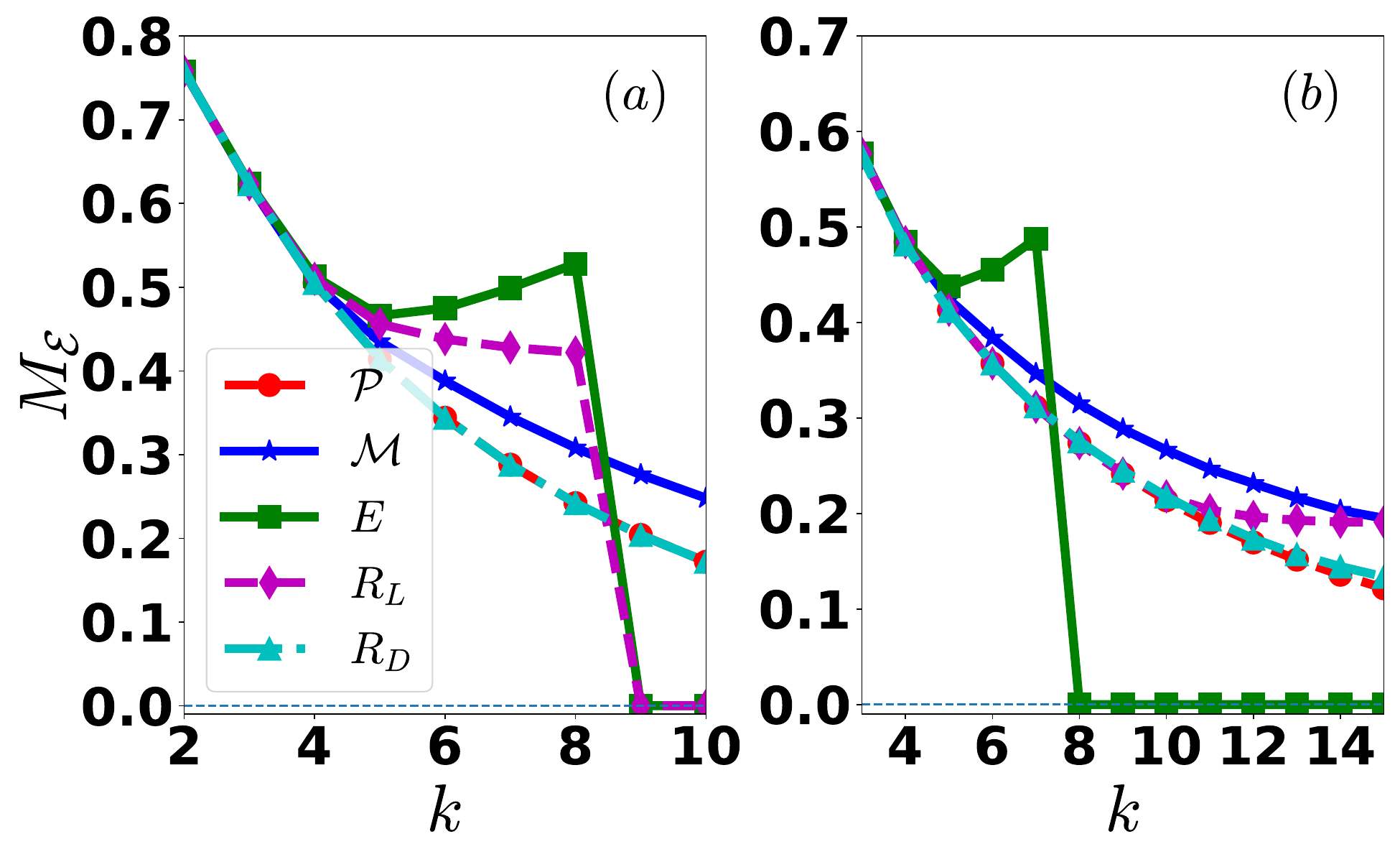}
    \captionsetup{justification=Justified,singlelinecheck=false}  \caption{The mean detectable entanglement, \(M_{\mathcal{E}}\) (vertical axis) vs the rank \(k\) (horizontal axis) of bipartite mixed states in two different local dimensions: (a) \(2 \otimes 5\) and (b) \(3 \otimes 5\). The entanglement detection criteria discussed in Eq.~(\ref{eq:measure_set}) are compared: \(\mathcal{P}\) (red circles), \(\mathcal{M}\) (blue stars), \(\mathrm{E}\) (green squares), \(\mathrm{R}_L\) (magenta diamonds), and \(\mathrm{R}_D\) (cyan triangles). Similar features are observed for low rank states although for moderate \(k\) values, the efficiency of \(\mathrm{R}_L\) and \(\mathrm{R}_D\) gets interchanged with the increase of \(d_1\).  Both axes are dimensionless.}
    \label{fig:mean_2_5_rank}
    \end{figure}

\begin{figure}
\includegraphics [width=\linewidth]{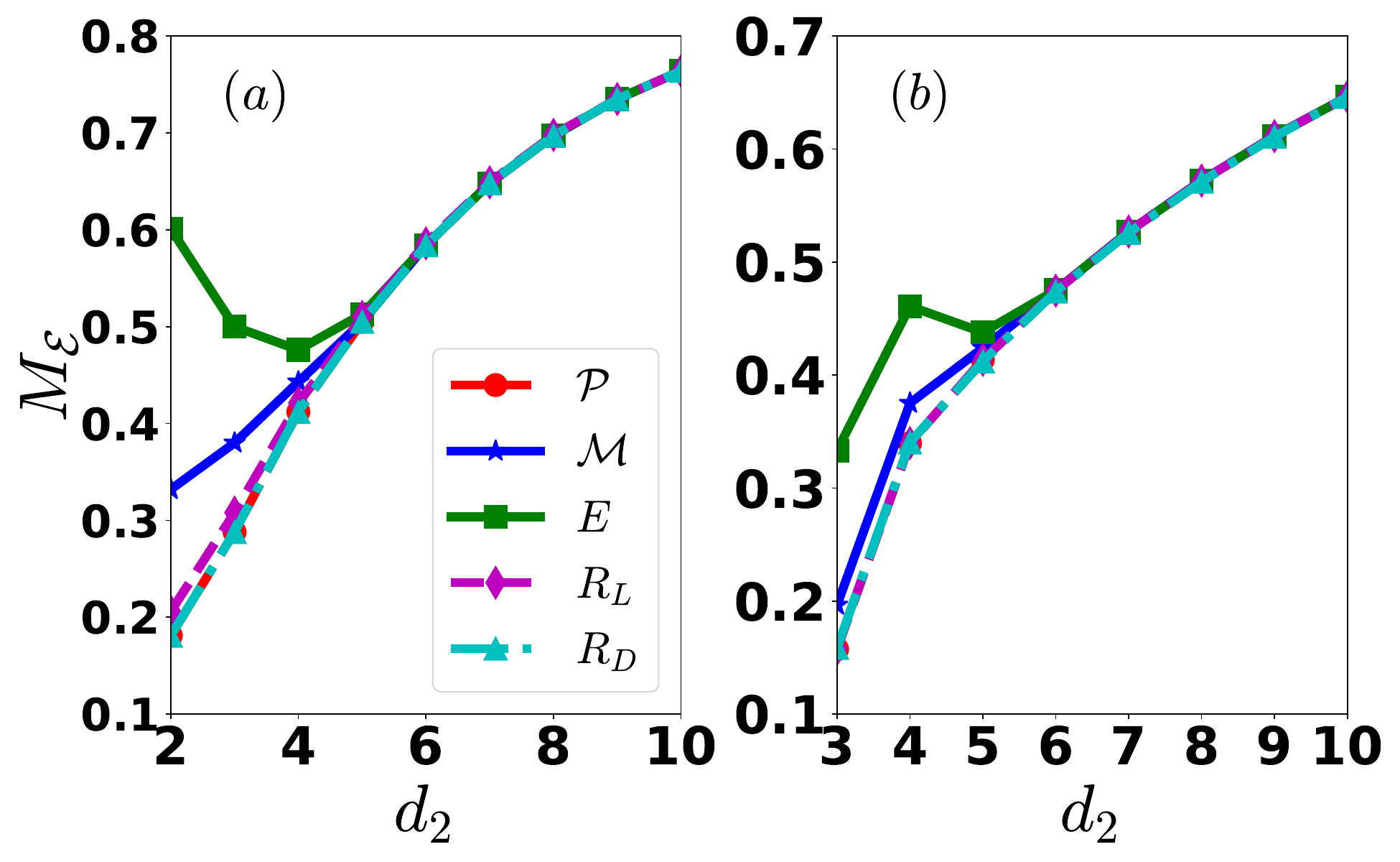}
\captionsetup{justification=Justified,singlelinecheck=false} \caption{Mean detectable entanglement, \(M_{\mathcal{E}}\) (vertical axis) with the subsystem dimension, \(d_2\) (horizontal axis) for states of rank (a) \(k = 4\) in \(2 \otimes d_2\), and (b) \(k = 5\) in \(3 \otimes d_2\). All other specifications are the same as in Fig.~\ref{fig:mean_2_5_rank}. Nonmonotonic behavior for entropic criterion is observed.  Both axes are dimensionless.}
  
\label{fig:mean_4_rank_d2}
\end{figure}

Let us now analyze the behaviors of \(M_{\mathcal{E}}\) and \(m_{\mathcal{E}}\) to understand whether the effectiveness of the criteria relies upon the average or minimum entanglement possessed by random states. Our investigation shows that indeed it is the case. Let us first elaborate it from the trends of \(M_{\mathcal{E}}\) -- $(1)$ Among Haar uniformly generated states, average entanglement \((\text{LN})\) decreases with the increase of the rank, \(k\), irrespective of the individual dimension (as shown in Fig. \ref{fig:mean_2_5_rank} for \(2\otimes 5\) and \(3\otimes 5\) systems). This may be attributed to the fact that as the rank increases, states with lower values of LN become predominant~\cite{Gupta_PRA_2022_non-Markovianity, Gupta_PRA_2022_catalysis}. $(2)$ In \(2\otimes d\) systems, the average \(M_{\mathrm{R}_D}\) takes the same path as \(M_{\mathcal{P}}\) since \(\text{LN}\) is based on the partial transposition criteria, $\mathcal{P}$, to which \(\mathrm{R}_D\) is equivalent in this regime~\cite{JBatle_2004_iop_allconnnection}, \aka{as we have also independently demonstrated in Proposition~\ref{pro:ppt_red_eqv}.} $(3)$ For small ranks, i.e., when \(k \leq \min[d_1,d_2]+2\), all entanglement detection methods behave similarly, i.e., the mean entanglement of typical states and mean entanglement detected by any criteria belonging to \(\mathcal{S}\) coincide.  $(4)$ On the other hand, for a fixed, \(k\) and \(d_1\), when the other subsystem dimension, \(d_2\), increases, \(M_{\mathcal{E}}\) increases for all identification methods (see Figs.~\ref{fig:mean_4_rank_d2} (a) and (b) for states in \(2\otimes d\)  with
$k = 4$ and in \(3\otimes d\) with \(k=5\) respectively), eluding to the fact that the entanglement of states increases with the dimension for fixed ranks. Further, by fixing \(d_1\) and \(k \lesssim d_2\)  \((d_2\geq d_1)\), the performance of all criteria becomes the same (see also Table~\ref{tab:frac_state_det_dim_new}). $(5)$ Interestingly, the values of \(M_{\mathcal{E}}\) associated with the entropic criterion, \(\mathrm{E}\), and the realignment criterion, \(\mathrm{R}_L\),
exhibits a non-monotonic trend with respect to \(k\), which may be attributed to the fact that with increasing rank, these two criteria become grossly ineffective in detecting low entangled states, i.e., $n/10^5 \ll 1$, near full rank, i.e., \(k = 9, 10\) (see Table.~\ref{tab:frac_state_detection}). In contrast, other criteria are successful in doing so. $(6)$ The mean entanglement, $M_\mathcal{M}$ corresponding to the majorization criterion, $\mathcal{M}$, is higher than that of the PT, \(\mathcal{P}\), and the reduction, \(\mathrm{R}_D\),  beyond low rank states although  \(M_{\mathrm{R}_L}\) turns out to be almost same with moderate rank (see Fig. \ref{fig:mean_2_5_rank}). This suggests that \(\mathcal{M}\) is unable to detect low entangled states as efficiently as the other two. Since increasing the rank also increases the likelihood of generating states with small entanglement content, the greater effectiveness of $\mathcal{P}$ and $\mathrm{R}_D$ becomes prominent at higher ranks.

The behavior of minimum entanglement, \(m_{\mathcal{E}}\), is similar to the mean, which again decreases with increasing rank, \(k\), for \(\mathcal{P}, \mathcal{M}, ~\text{and}~ \mathrm{R}_D\). 
A non-monotonic behavior emerges for \(m_{\mathrm{E}}\) and \(m_{\mathrm{R}_L}\)(see Tables.~\ref{tab:emin_table_3_5} and ~\ref{tab:emin_fixed_k5}), thereby showing that $\mathrm{E}$ and $\mathrm{R}_L$ decrease in efficiency with increasing rank, because they cannot detect low entangled states. Both the nature of \(M_{\mathcal{E}}\) and \(m_\mathcal{E}\) establish that when the entanglement content of a typical state is high, which surely occurs for low rank states, all the existing entanglement detection methods are equally effective. 
Since, after a certain rank, $m_\mathcal{E}$ is higher for the reduction criterion as compared to the PPT criterion, it implies that the reduction criterion is unable to detect low entangled states in $d_1 \otimes d_2$ when $d_1, d_2 > 2$. Thus, we can conclude that according to the efficacy in detecting entangled states, $\mathrm{R}_D$ is a strict subset of $\mathcal{P}$. The realignment criterion follows PPT and reduction, while the majorization and the entropy criteria fare the worst. This is vindicated by the fact that $m_\mathcal{M} > m_{\mathrm{R}_L} > m_{\mathrm{R}_D} > m_\mathcal{P}$, while the entropy criterion is unable to detect entangled states beyond rank $k > 7$ in $3 \otimes 5$ (see Table.~\ref{tab:emin_table_3_5}, and also supported by Proposition.~\ref{pro:entropy_full_rank}) and also has the highest minimum detectable entanglement for $k = 5$ states in $3 \otimes d$, as elucidated in Table.~\ref{tab:emin_fixed_k5}.
 Finally, the realignment criterion is effective for low rank \(d_1\otimes d_2\) states (\(d_1>2, d_2>3\)). These results are in good agreement with the ones obtained by studying the fraction of states.

\textbf{Remark $\mathbf{2}$.} \aka{The non-monotonicity exhibited in $M_\mathcal{E}$ by the entropy criterion can be explained through its inherent inefficiency in detecting entanglement of high-rank states. As shown in Proposition \ref{pro:entropy_full_rank}}, it cannot detect entangled states with rank $k > \max(d_1, d_2)$. Therefore, for such states, both the mean and minimum reduce to zero abruptly as soon as the critical rank is reached. Furthermore, Table.~\ref{tab:frac_state_detection} elucidates how the fraction of detected states decreases sharply with increasing rank, as a result of which, for ranks close to $\max(d_1, d_2)$, the mean and minimum detected entanglement shoot up for the entropy criterion because the denominator in the definition of $M_\mathcal{E}$ and $m_\mathcal{E}$ (Eq.\eqref{eq:min_ent}) assumes very low values. This results in the non-monotonic behavior of the figures of merit for the entropy criterion.

\begin{table}[H]
\centering
\renewcommand{\arraystretch}{1.2}
\begin{tabular}{|>{\centering\arraybackslash}p{1.2cm}|>{\centering\arraybackslash}p{1.3cm}|>{\centering\arraybackslash}p{1.3cm}|>{\centering\arraybackslash}p{1.3cm}|>{\centering\arraybackslash}p{1.3cm}|>{\centering\arraybackslash}p{1.3cm}|}
\hline
\textbf{\(d_1 \otimes d_2\)} & \textbf{\(M_\mathcal{P}\)} & \textbf{\(M_\mathcal{M}\)} & \textbf{\(M_E\)} & \textbf{\(M_{\mathrm{R}_L}\)} & \textbf{\(M_{\mathrm{R}_D}\)} \\
\hhline{|======|}
\(3 \otimes 3\)  & \(0.138\) & \(0.237\) & \textemdash & \(0.145\) & \(0.163\) \\ \hline
\(3 \otimes 4\)  & \(0.211\) & \(0.275\) & \textemdash & \(0.211\) & \(0.212\) \\ \hline
\(3 \otimes 5\)  & \(0.274\) & \(0.315\) & \textemdash & \(0.274\) & \(0.274\) \\ \hline
\(3 \otimes 6\)  & \(0.329\) & \(0.355\) & \textemdash & \(0.329\) & \(0.329\) \\ \hline
\(3 \otimes 7\)  & \(0.377\) & \(0.394\) & \(0.448\) & \(0.377\) & \(0.377\) \\ \hline
\(3 \otimes 8\)  & \(0.420\) & \(0.425\) & \(0.435\) & \(0.420\) & \(0.420\) \\ \hline
\(3 \otimes 9\)  & \(0.458\) & \(0.458\) & \(0.458\) & \(0.458\) & \(0.458\) \\ \hline
\(3 \otimes 10\) & \(0.493\) & \(0.493\) & \(0.493\) & \(0.493\) & \(0.493\) \\ \hline
\end{tabular}
\captionsetup{justification=Justified,singlelinecheck=false} 
\caption{The mean value of each criterion, \(\mathcal{E}\), for states of rank \(k = 8\) in \(3 \otimes d_2\). A dash (\textemdash) denotes no detection of entanglement.}
\label{tab:frac_state_det_dim_new}
\end{table}

\begin{table}[H]
\centering
\renewcommand{\arraystretch}{1.2}
\begin{tabular}{|>{\centering\arraybackslash}p{1.2cm}|>{\centering\arraybackslash}p{1.3cm}|>{\centering\arraybackslash}p{1.3cm}|>{\centering\arraybackslash}p{1.3cm}|>{\centering\arraybackslash}p{1.3cm}|>{\centering\arraybackslash}p{1.3cm}|}
\hline
\textbf{\(k\)} & \textbf{\(\mathcal{P}\)} & \textbf{\(\mathcal{M}\)} & \textbf{\(\mathrm{E}\)} & \textbf{\(\mathrm{R}_L\)} & \textbf{\(\mathrm{R}_D\)} \\
\hhline{|======|}
\(2\)  & \(0.404\) & \(0.404\) & \(0.404\) & \(0.404\) & \(0.404\) \\ \hline
\(3\)  & \(0.386\) & \(0.386\) & \(0.386\) & \(0.386\) & \(0.386\) \\ \hline
\(4\)  & \(0.320\) & \(0.320\) & \(0.340\) & \(0.320\) & \(0.320\) \\ \hline
\(5\)  & \(0.252\) & \(0.298\) & \(0.334\) & \(0.252\) & \(0.252\) \\ \hline
\(6\)  & \(0.220\) & \(0.272\) & \(0.379\) & \(0.220\) & \(0.220\) \\ \hline
\(7\)  & \(0.172\) & \(0.229\) & \(0.488\) & \(0.172\) & \(0.172\) \\ \hline
\(8\)  & \(0.149\) & \(0.205\) & \textemdash & \(0.149\) & \(0.149\) \\ \hline
\(9\)  & \(0.117\) & \(0.187\) & \textemdash & \(0.133\) & \(0.117\) \\ \hline
\(10\) & \(0.086\) & \(0.178\) & \textemdash & \(0.121\) & \(0.089\) \\ \hline
\(11\) & \(0.080\) & \(0.168\) & \textemdash & \(0.129\) & \(0.087\) \\ \hline
\(12\) & \(0.058\) & \(0.160\) & \textemdash & \(0.127\) & \(0.072\) \\ \hline
\(13\) & \(0.046\) & \(0.145\) & \textemdash & \(0.123\) & \(0.069\) \\ \hline
\(14\) & \(0.038\) & \(0.145\) & \textemdash & \(0.137\) & \(0.055\) \\ \hline
\(15\) & \(0.025\) & \(0.156\) & \textemdash & \(0.143\) & \(0.053\) \\ \hline
\end{tabular}
\captionsetup{justification=Justified,singlelinecheck=false} 
\caption{The minimum entanglement, \(m_{\mathcal{E}}\), detected by various criteria in the set \(\mathcal{S}\), is presented for a fixed total system dimension of \(3 \otimes 5\) by varying rank \(2 \leq k \leq 15\). All other specifications are the same as in Table~\ref{tab:frac_state_det_dim_new}.}
\label{tab:emin_table_3_5}
\end{table}

\begin{table}[h!]
\centering
\renewcommand{\arraystretch}{1.2}
\begin{tabular}{|>{\centering\arraybackslash}p{1.2cm}|>{\centering\arraybackslash}p{1.3cm}|>{\centering\arraybackslash}p{1.3cm}|>{\centering\arraybackslash}p{1.3cm}|>{\centering\arraybackslash}p{1.3cm}|>{\centering\arraybackslash}p{1.3cm}|}
\hline
\textbf{\(d_1 \otimes d_2\)} & \textbf{\(\mathcal{P}\)} & \textbf{\(\mathcal{M}\)} & \textbf{\(\mathrm{E}\)} & \textbf{\(\mathrm{R}_L\)} & \textbf{\(\mathrm{R}_D\)} \\
\hhline{|======|}
\(3 \otimes 3\)   & $0.036$ & $0.104$ & $0.261$ & $0.036$ & $0.042$ \\ \hline
\(3 \otimes 4\)   & $0.168$ & $0.241$ & $0.371$ & $0.168$ & $0.168$ \\ \hline
\(3 \otimes 5\)   & $0.252$ & $0.298$ & $0.333$ & $0.252$ & $0.252$ \\ \hline
\(3 \otimes 6\)   & $0.311$ & $0.311$ & $0.355$ & $0.311$ & $0.311$ \\ \hline
\(3 \otimes 7\)   & $0.381$ & $0.381$ & $0.403$ & $0.381$ & $0.381$ \\ \hline
\(3 \otimes 8\)   & $0.460$ & $0.460$ & $0.460$ & $0.460$ & $0.460$ \\ \hline
\(3 \otimes 9\)   & $0.501$ & $0.501$ & $0.501$ & $0.501$ & $0.501$ \\ \hline
\(3 \otimes 10\)  & $0.544$ & $0.544$ & $0.544$ & $0.544$ & $0.544$ \\ \hline
\end{tabular}
\captionsetup{justification=Justified,singlelinecheck=false} 
\caption{The minimum entanglement, \(m_{\mathcal{E}}\), detected by various criteria is presented for a fixed rank \(k=5\) for varying subsystem dimension, \(d_2\), of states in \(3 \otimes d_2\). All other specifications are the same as in Table~\ref{tab:frac_state_det_dim_new}.}
\label{tab:emin_fixed_k5}
\end{table}

\textcolor{black}{
\subsubsection{Numerical convergence of the figures of merit}
\label{app:convergence}
The entire analysis in this paper is presented on a sample of $10^{6}$ randomly generated states. To ensure the convergence of the numerical results, we perform additional simulations with samples of different sizes. Specifically, we compute all the figures of merit, viz, the normalized fraction $F^\mathcal{E}$, the mean detected entanglement $M_\mathcal{E}$, and the minimum detected $m_\mathcal{E}$ for samples comprising $10^4, 5 \times 10^4, 10^5, 3 \times 10^5, 5 \times 10^5, 8 \times 10^5 ~\text{and}~ 10^6$ Haar random states, and observe that the last three sets provide the same values of the aforementioned quantities up to the third decimal place, thereby confirming the convergence of the data
. As an instance, we enlist the three figures of merit corresponding to the reduction criterion for states in $3 \otimes 5$ of ranks $2$ and $15$ in Table.~\ref{tab:convergence}. A similar observation is ensured for all other detection criteria by our numerical simulations.
\\
\begin{table}[H]
\centering
\renewcommand{\arraystretch}{1.3}
\begin{tabular}{|c|c|c|c|c|}
\hline
\textbf{Sample size} & \textbf{$k$} & $F^{\mathcal{E}}$ & $ M_\mathcal{E}$ & $ m_\mathcal{E}$ \\
\hline
\multirow{2}{*}{$1\times10^4$}
& $2$  & $1$ & $0.702(5)$ & $0.499(9)$  \\
& $15$ & $0.566(1)$ & $0.133(7)$ & $0.069(3)$\\
\hline
\multirow{2}{*}{$5\times10^4$}
& $2$  & $1$ & $0.701(5)$ & $0.493(9)$\\
& $15$ & $0.575(1)$ &  $0.133(8)$ &$0.064(6)$ \\
\hline
\multirow{2}{*}{$1\times10^5$}
& $2$  & $1$ & $0.701(7)$ & $0.455(7)$ \\
& $15$ & $0.572(4)$ & $0.133(7)$ & $0.058(1)$ \\
\hline
\multirow{2}{*}{$3\times10^5$}
& $2$  & $1$ & $0.701(5)$ & $0.438(5)$ \\
& $15$ & $0.572(9)$ & $0.133(7)$ & $0.058(8)$ \\
\hline
\multirow{2}{*}{$5\times10^5$}
& $2$  & $1$ & $0.701(5)$ &  $0.452(0)$    \\
& $15$ & $0.573(7)$ & $0.133(6)$ & $0.053(3)$     \\
\hline
\multirow{2}{*}{$8\times10^5$}
& $2$  & $1$ & $0.701(5)$ &  $0.428(9)$    \\
& $15$ & $0.573(1)$ & $0.133(6)$ & $0.060(1)$     \\
\hline
\multirow{2}{*}{$1\times10^6$}
& $2$  & $1$ & $0.701(6)$ &  $0.427(2)$    \\
& $15$ & $0.573(1)$ & $0.133(7)$ & $0.055(5)$     \\
\hline
\end{tabular}
\caption{Mean value $ M_{\mathcal{E}}$, minimum value $m_{\mathcal{E}}$, and normalized fraction $F^{\mathcal{E}}$ for $\mathcal{E} = \mathrm{R}_D$ (reduction criterion), computed over different sample sizes and ranks $k$. To argue convergence up to the third decimal place, the numbers in $()$ indicate the value of the corresponding quantity at the fourth decimal place, showing that any change in the numerical results occurs beyond the third decimal place for the sample sizes considered in the manuscript. The above table is for system $3 \otimes 5$. The results demonstrate the stability and convergence of the normalized fraction and mean with increasing sample size up to the third decimal place. Since minimum values correspond to extreme events in the distribution, observing the same minimum value repeatedly is unlikely, i.e., generating the state with the same minimum entanglement across all sample sizes is statistically not possible, and thus we do not expect any convergence in $m_{\mathcal{E}}$ beyond the first decimal place.}
\label{tab:convergence}
\end{table}
}

\subsection{Crucial role of asymmetry in subsystem's dimension on detection efficiency}
\label{subsec:mean_min_same_dimension}

For pure states in $d_1 \otimes d_2$, the entanglement content is upper-bounded by the von Neumann entropy of the smaller subsystem dimension~\cite{RevModPhys_2009_horo}, say $d_1$, due to the Schmidt decomposition. As a result, if, keeping the smaller subsystem dimension fixed, the total dimension of the system increases, the entanglement content, and thereby the mean and minimum detectable entanglement are expected to remain the same with an increase in the asymmetry, $|d_1 - d_2|$, in the two subsystem dimensions. However, the situation is not straightforward in the case of mixed states. Our goal in this section is to understand {\it how the effectiveness of the different entanglement detection criteria changes with \(|d_1-d_2|\) for bipartite mixed states}. While the entanglement content of states is known to increase with the total dimension~\cite{Gupta_PRA_2022_catalysis}, its dependence on the asymmetry, independent of the total dimension, has not been studied in the literature. Moreover, since we quantify entanglement using logarithmic negativity (LN), which is only known to upper-bound distillable entanglement and does not have a simple relation with local subsystem entropy, intuition from pure states does not directly apply. We will show how the variation of entanglement with subsystem asymmetry directly affects the performance of different detection criteria, particularly the realignment criterion when moving beyond qubit systems. To the best of our knowledge, such a study on the dependence of a criterion's detection efficiency (beyond its role of quantifying the entanglement) has not been reported earlier. Thus, while we have so far concentrated on the performance of the various criteria under the conditions of varying rank and system dimension, it is worthwhile to investigate whether the difference in the subsystem dimensions also plays a role in a criterion's efficacy. Specifically, fixing the total dimension to $d_{12} = d_1 d_2$, we wish to analyze how the mean detected entanglement behaves with $|d_1 - d_2|$. While it is apparent that an increase in $d_{12}$ leads to states with greater entanglement content, and hence the criteria retain their effectiveness even at higher ranks, the dependence on $|d_1 - d_2|$ is not clearly studied for typical states in the literature. In particular, for a given \(d_{12}\) and rank, especially the lowest and highest ranks, we study the variation of $M_\mathcal{E}$, and $m_\mathcal{E}$ for Haar uniform states.

Let us consider the smallest total dimension expressible as different products of two distinct integers, i.e., \(12\). It can be considered as \(2 \otimes 6\) and \(3 \otimes 4\). We observe that the mean entanglement, \({M}_{\mathcal{E}}\), and the minimum, $m_\mathcal{E}$, for all the measures contained in the set \(\mathcal{S}\) are higher when the difference between subsystem dimensions is greater, for a fixed rank, $k = 2$ (lowest rank beyond pure states) or maximum rank, $12$. This may indicate that more entangled states are generated on average as \(|d_1-d_2|\) grows for a fixed \(d_{12}\) and \(k\), and hence average entanglement required to determine the entanglement via a criterion is also high.
This trend persists in higher dimensions; for example, from \(d_{12} =18\) which can be considered as \(2 \otimes 9\) and \(3 \otimes 6\), to $d_{12} = 36\) as $2 \otimes 18, 3\otimes 12, 4 \otimes 9,$ and $6 \otimes 6$, a greater imbalance in the subsystem sizes again leads to stronger mean and minimum entanglement (see Table.~\ref{tab:asymmetry}).
For rank-$2$ states across systems with varying local dimensions, all the entanglement detection criteria exhibit equivalent strength. In contrast, for full-rank states, the hierarchy follows the structure as discussed before, i.e., only \(\mathrm{R}_D\), \(\mathcal{M}\), and sometimes \(\mathrm{R}_L\) become effective.  Interestingly, we notice that the realignment criterion \(\mathrm{R}_L\) becomes adequate, only when the absolute difference in subsystem dimensions \(|d_1 - d_2|\) is small - case in point, when $d_1 - d_2 \leq 1$, $M_{\mathrm{R}_L} > 0$ for $d_{12} = 12, 16,$ and $36$ in Table.~\ref{tab:asymmetry} - otherwise, its detection capability vanishes. This was also vindicated in the analysis on the normalised fraction of detected states, specifically Table.~\ref{tab:sub_dim_and_rank_vary}. An intuition into this rise in detection ability for $\mathrm{R}_L$ can be obtained from Proposition \eqref{pro:real_detec}. For $d_2 > d_1 \gg 2$, and $d_2 - d_1 = 1$, we must have the rank satisfy $k \gtrsim d_1^2 d_2$ for realignment to be unsuccessful. However, since $k \in (2, d_1 d_2)$, the condition does not hold for small dimensional asymmetry. We can conclude that the entanglement content of a state and the detection efficiency of a given criterion not only depend on the rank and dimension, but also on the difference in subsystem dimensions, i.e., on how the state is constituted.

We can try to explain the increase in mean (minimum) entanglement, $M_\mathcal{E} (m_\mathcal{E})$, with higher differences in subsystem dimension, $|d_1 - d_2|$, for rank $k = 2$ as follows. Note that the total subsystem dimension, $d_{12} \gg k$, for rank-$2$ states under consideration. As a result the entropy of the total state, in $d_1 \otimes d_2$, relative to the maximal achievable average entropy is given as, $\Big(\log 2 - 1/d_{12} \Big)/\log d_{12} \to 0$ as $d_{12}$ increases~\cite{Kendon2002_pra}. Thus, with increasing $d_{12}$, the bipartite state moves further and further away from the maximally mixed state of its corresponding dimension. Furthermore, the purity of the rank-$2$ state, $(d_{12} + 2)/(1 + 2 d_{12}) \to 1/2$ for $d_{12} \gg 2$~\cite{Kendon2002_pra}. This again shows that the $d_{12}$-dimensional state has purity equal to the maximally mixed state of a dimension much lower than its own. Therefore, effectively, we can consider the state to be a pure state for a large total dimension, $d_{12}$. In that case, its entanglement content is given by the von Neumann entropy of its reduced density matrices, $\log d_1 - d_1/2 d_2$~\cite{Kendon2002_pra}, assuming $d_2 \geq d_1$ without loss of generality. In the two extreme cases, $d_1 = d_2 ~\text{or}~ (d_2 - d_1) = 0$ and $d_1 \ll d_2 ~\text{or}~ (d_2 - d_1) \gg 1$, we have the entanglement approximately $\log d_1 - 1/2$ and $\log d_1$ respectively. Thus, this serves as an intuitive argument for the increase in mean detected entanglement of the rank-$2$ state with higher differences in subsystem dimensions.

\begin{widetext}
\begin{minipage}{\textwidth}
\begin{table}[H]
\centering
\renewcommand{\arraystretch}{0.4}
\begin{tabular}{|>{\centering\arraybackslash}p{1.2cm}|
                >{\centering\arraybackslash}p{1.2cm}|
                >{\centering\arraybackslash}p{2.0cm}|
                >{\centering\arraybackslash}p{2.0cm}|
                >{\centering\arraybackslash}p{2.0cm}|
                >{\centering\arraybackslash}p{2.0cm}|
                >{\centering\arraybackslash}p{2.0cm}|
                >{\centering\arraybackslash}p{2.0cm}|}
\hline
\textbf{\(d_{12}\)} & {\(d_1\otimes d_2\)} & {\(k\)} & \textbf{$\mathcal{P}$} & \textbf{$\mathcal{M}$} & \textbf{$\mathrm{E}$} & \textbf{$\mathrm{R}_L$} & \textbf{$\mathrm{R}_D$} \\
\hline\hline
\multirow{4}{*}{$12$} 
& \multirow{2}{*}{$2 \otimes 6$} & $2$  & $0.803(0.363)$ & $\leftrightarrow$ & $\leftrightarrow$ & $\leftrightarrow$ & $\leftrightarrow$ \\ \cline{3-8}
&                               & $12$ & $0.174(0.003)$ & $0.236(0.095)$ & \textemdash & \textemdash & $0.174(0.003)$ \\ \cline{2-8}
& \multirow{2}{*}{$3 \otimes 4$} & $2$  & $0.625(0.329)$ & $\leftrightarrow$ & $\leftrightarrow$ & $\leftrightarrow$ & $\leftrightarrow$ \\ \cline{3-8}
&                               & $12$ & $0.120(0.005)$ & $0.207(0.138)$ & \textemdash & $0.149(0.073)$ & $0.141(0.052)$ \\ \hline

\multirow{4}{*}{$16$} 
& \multirow{2}{*}{$2 \otimes 8$} & $2$  & $0.855(0.495)$ & $\leftrightarrow$ & $\leftrightarrow$ & $\leftrightarrow$ & $\leftrightarrow$ \\ \cline{3-8}
&                               & $16$ & $0.176(0.028)$ & $0.221(0.108)$ & \textemdash & \textemdash & $0.176(0.028)$ \\ \cline{2-8}
& \multirow{2}{*}{$4 \otimes 4$} & $2$  & $0.597(0.392)$ & $\leftrightarrow$ & $\leftrightarrow$ & $\leftrightarrow$ & $\leftrightarrow$ \\ \cline{3-8}
&                               & $16$ & $0.098(0.026)$ & \textemdash & \textemdash & $0.106(0.054)$ & $0.148(0.099)$ \\ \hline

\multirow{4}{*}{$18$} 
& \multirow{2}{*}{$2 \otimes 9$} & $2$  & $0.873(0.528)$ & $\leftrightarrow$ & $\leftrightarrow$ & $\leftrightarrow$ & $\leftrightarrow$ \\ \cline{3-8}
&                               & $18$ & $0.176(0.055)$ & $0.216(0.133)$ & \textemdash & \textemdash & $0.176(0.055)$ \\ \cline{2-8}
& \multirow{2}{*}{$3 \otimes 6$} & $2$  & $0.763(0.513)$ & $\leftrightarrow$ & $\leftrightarrow$ & $\leftrightarrow$ & $\leftrightarrow$ \\ \cline{3-8}
&                               & $18$ & $0.123(0.039)$ & $0.185(0.145)$ & \textemdash & \textemdash & $0.130(0.059)$ \\ \hline

\multirow{6}{*}{$24$} 
& \multirow{2}{*}{$2 \otimes 12$} & $2$  & $0.906(0.652)$ & $\leftrightarrow$ & $\leftrightarrow$ & $\leftrightarrow$ & $\leftrightarrow$ \\ \cline{3-8}
&                               & $24$ & $0.177(0.081)$ & $0.206(0.132)$ & \textemdash & \textemdash & $0.177(0.081)$ \\ \cline{2-8}
& \multirow{2}{*}{$3 \otimes 8$} & $2$  & $0.830(0.640)$ & $\leftrightarrow$ & $\leftrightarrow$ & $\leftrightarrow$ & $\leftrightarrow$ \\ \cline{3-8}
&                               & $24$ & $0.124(0.050)$ & $0.171(0.152)$ & \textemdash & \textemdash & $0.127(0.070)$ \\ \cline{2-8}
& \multirow{2}{*}{$4 \otimes 6$} & $2$  & $0.720(0.547)$ & $\leftrightarrow$ & $\leftrightarrow$ & $\leftrightarrow$ & $\leftrightarrow$ \\ \cline{3-8}
&                               & $24$ & $0.101(0.040)$ & $0.145(0.121)$ & \textemdash & \textemdash & $0.132(0.056)$ \\ \hline

\multirow{8}{*}{$36$} 
& \multirow{2}{*}{$2 \otimes 18$} & $2$  & $0.938(0.735)$ & $\leftrightarrow$ & $\leftrightarrow$ & $\leftrightarrow$ & $\leftrightarrow$ \\ \cline{3-8}
&                               & $36$ & $0.177(0.109)$ & $0.196(0.139)$ & \textemdash & \textemdash & $0.177(0.109)$ \\ \cline{2-8}
& \multirow{2}{*}{$3 \otimes 12$} & $2$  & $0.891(0.751)$ & $\leftrightarrow$ & $\leftrightarrow$ & $\leftrightarrow$ & $\leftrightarrow$ \\ \cline{3-8}
&                               & $36$ & $0.125(0.063)$ & \textemdash & \textemdash & \textemdash & $0.127(0.066)$ \\ \cline{2-8}
& \multirow{2}{*}{$4 \otimes 9$} & $2$  & $0.827(0.695)$ & $\leftrightarrow$ & $\leftrightarrow$ & $\leftrightarrow$ & $\leftrightarrow$ \\ \cline{3-8}
&                               & $36$ & $0.102(0.068)$ & \textemdash & \textemdash & \textemdash & $0.121(0.103)$ \\ \cline{2-8}
& \multirow{2}{*}{$6 \otimes 6$} & $2$  & $0.674(0.576)$ & $\leftrightarrow$ & $\leftrightarrow$ & $\leftrightarrow$ & $\leftrightarrow$ \\ \cline{3-8}
&                               & $36$ & $0.080(0.052)$ & \textemdash & \textemdash & $0.090(0.184)$ & \textemdash \\ \hline
\end{tabular}
\caption{The mean (minimum) detectable entanglement, \(M_\mathcal{E} (m_\mathcal{E})\), for various entanglement criterion (columns),  considered in Eq.~(\ref{eq:measure_set}) is evaluated for fixed total system dimensions \(d_{12} = 12, 16, 18, 24, 36\), where the subsystem dimensions \(d_1, d_2\) are varied such that the product \(d_1 d_2 = d_{12}\). We consider two extreme cases of ranks, \(k = 2\) and \(k = d_1 d_2\). The $\leftrightarrow$ denotes that the values in that row are the same as those of $\mathcal{P}$. All other specifications are the same as in Table.~\ref{tab:frac_state_det_dim_new}. Interestingly, the results display that the performance of the entanglement criteria depends on the difference between local dimensions, \(|d_1-d_2|\). }
\label{tab:asymmetry}
\end{table}
\end{minipage}
\end{widetext}

\section{Conclusion}
\label{sec:conclusion}

Entanglement, the earliest and most extensively studied form of quantum correlation, continues to play a central role in the development of quantum information theory. 
Although the benefits of entangled states are well-established across a range of quantum technologies, 
the rich structure of entangled states in arbitrary dimensions, both bipartite and multipartite, is yet to be fully understood.
One of the central questions in the theory of entanglement is how to definitively classify states as entangled or separable, thereby identifying their usefulness for information-theoretic applications. While several detection methods have been developed, their computation and experimental implementation remain a significant challenge. On the other hand, criteria that can be efficiently calculated and realized are not always necessary and sufficient for detecting usable entanglement. Thus, there is a strong need for a systematic classification of entanglement detection criteria, alongside a rigorous evaluation of their effectiveness, especially for randomly generated quantum states. While the entanglement properties of generic multiqubit states have been explored to some extent, the question of whether random quantum states in higher dimensions exhibit any form of universal entanglement behavior remains unresolved. Addressing this issue could not only deepen our theoretical understanding but also guide the development of more scalable and experimentally viable entanglement detection techniques.

We analyzed the performance of four widely used entanglement detection criteria - majorization, reduction, realignment, and entropy-based methods - by benchmarking them against the partial transposition (PT) criterion and the PT-based logarithmic negativity measure. Focusing on bipartite quantum states with non-positive partial transpose (NPT), we examined the detection efficacy of the various criteria in terms of three factors -- the dependence on the total dimension of the states, the effect of rank, and the difference of local dimensions -- in ascertaining entanglement of random mixed states in arbitrary dimensions. This comprehensive approach enabled us to construct a mathematically robust framework for assessing the universality and reliability of these detection methods in arbitrary finite dimensions. In particular, we provided analytical lower bounds on the rank of quantum states beyond which the realignment and entropy-based criteria fail to detect entanglement, thus highlighting intrinsic limitations of these approaches in identifying high-rank entangled states in high-dimensional systems.

For studying features of random states, we defined three figures of merit  -- the fraction of detected states, the mean detectable entanglement, and the minimum entanglement required for detection. The fraction of detected states offered a straightforward metric, reflecting how effectively a given criterion can identify entangled states among Haar-randomly simulated samples — the higher this fraction, the more powerful the criterion.  In contrast, the mean and minimum detectable entanglement provided deeper insight into the entanglement ``threshold'' required for a criterion to function. These metrics helped to quantify the entanglement necessary for detection and established practical bounds on the resources each criterion needed to reliably identify entangled states across various dimensions and ranks.

We revealed a distinct hierarchy of entanglement detection criteria in qudit-qudit systems, which differs notably from the behavior observed in qubit-qudit systems. In particular, we found that the realignment criterion consistently outperforms the majorization criterion in systems beyond \(2\otimes d\).
During the course of our study, we provided independent proof of the equivalence of the PT and reduction
criteria in qubit-qudit systems.  Our results showed that the fraction of detected entangled states decreases with increasing rank but improves with the overall system dimension. The  hierarchy based on the fraction of detected states remains consistent  when the strength of the
criteria is judged through the quantitative lens of the
mean and minimum entanglement. While the entropic criterion has clear operational relevance in quantum communication, it was the least effective across all systems studied. Conversely, although the realignment criterion can detect positive partially transposed bound entangled states, it loses effectiveness significantly as the rank of NPT states increases, revealing its limitations in high-rank scenarios.

The observed hierarchies among entanglement detection methods for generic states open up intriguing avenues for exploring similar structures in other resource theories, such as non-locality and non-stabilizerness. Additionally, assessing these criteria based on their ability to predict quantum advantage in computational and information-theoretic tasks is a promising direction.

\section*{Acknowledgment}
The authors acknowledge the cluster computing facility at Harish-Chandra Research Institute, together with the use of Armadillo~\cite{Sanderson_JOSS_2016_armadillo, Sanderson_2019_MCA_armadillo}, and QIClib -- a modern C++ library for general-purpose quantum information processing and quantum computing~\cite{QICLib}. A.K.A., S.M., and A.S.D. acknowledge support from the project entitled ``Technology Vertical - Quantum Communication'' under the National Quantum Mission of the Department of Science and Technology (DST)  (Sanction Order No. DST/QTC/NQM/QComm/$2024/2$ (G)). R.G. acknowledges funding from the HORIZON-EIC-$2022$-PATHFINDERCHALLENGES-$01$ program under Grant Agreement No.~$10111489$ (Veriqub). Views and opinions expressed are, however, those of the authors only and do not necessarily reflect those of the European Union. Neither the European Union nor the granting authority can be held responsible for them.

\bibliography{ref}
\end{document}